\documentclass[a4paper,fleqn,usenatbib]{mnras}

\usepackage{newtxtext,newtxmath}
\usepackage{float}

\usepackage[T1]{fontenc}
\usepackage{ae,aecompl}

\usepackage{graphicx}	
\usepackage{amsmath}	
\usepackage{amssymb}	
\usepackage{booktabs}
\usepackage[flushleft]{threeparttable}

\title[Star Formation History of WLM]{Star formation at the edge of the Local Group: a rising star formation history in the isolated galaxy WLM\thanks{Based on observations made with the NASA/ESA
    {\it Hubble Space Telescope}, obtained at the Space Telescope Science
    Institute, which is operated by the Association of Universities for
    Research in Astronomy, Inc., under NASA contract NAS5-26555. These
    observations are associated with program 13768.}}

\author[Albers et al.]{Saundra M. Albers,$^{1}$\thanks{E-mail: saundraalbers@berkeley.edu}
Daniel R. Weisz,$^{1}$\thanks{E-mail: dan.weisz@berkeley.edu}
Andrew A. Cole,$^{2}$
Andrew E. Dolphin,$^{3}$ \newauthor
Evan D. Skillman,$^{4}$
Benjamin F. Williams,$^{5}$
Michael Boylan-Kolchin,$^{6}$\newauthor
James S. Bullock,$^{7}$
Julianne J. Dalcanton,$^{5}$
Philip F. Hopkins,$^{8}$
Ryan Leaman,$^{9}$ \newauthor
Alan W. McConnachie,$^{10}$
Mark Vogelsberger,$^{11}$
Andrew Wetzel$^{12}$
\\
$^{1}$Department of Astronomy, University of California Berkeley, Berkeley, CA 94720 \\
$^{2}$School of Natural Sciences, University of Tasmania, Private Bag 37, Hobart, 
Tasmania, 7001 Australia \\
$^{3}$Raytheon, 1151 E. Hermans Rd, Tucson, AZ 85756, USA \\
$^{4}$Minnesota Institute for Astrophysics, University of Minnesota, Minneapolis, MN 55455, USA \\
$^{5}$Department of Astronomy, Box 351580, University of Washington, Seattle, WA 98195, USA \\
$^{6}$Department of Astronomy, The University of Texas at Austin, Austin, TX 78712, USA \\
$^{7}$Center for Cosmology, Department of Physics \& Astronomy, University of California, Irvine, 4129 Reines Hall, Irvine, CA 92697, USA \\
$^{8}$TAPIR, Mailcode 350-17, California Institute of Technology, Pasadena, CA 91125, USA \\
$^{9}$Max-Planck Institut f{\"u}r Astronomie, K{\"o}nigstuhl 17, D-69117 Heidelberg, Germany \\
$^{10}$NRC Herzberg Astronomy and Astrophysics, 5071 West Saanich Road, Victoria, BC V9E 2E7, Canada \\
$^{11}$Kavli Institute for Astrophysics and Space Research, Massachusetts Institute of Technology, Cambridge, MA 02139 \\
$^{12}$Department of Physics, University of California, Davis, CA 95616, USA \\
}

\date{Submitted June 21 2019}

\pubyear{2019}

\begin{document}
\label{firstpage}
\pagerange{\pageref{firstpage}--\pageref{lastpage}}
\maketitle

\begin{abstract}
We present the star formation history (SFH) of the isolated ($D\sim 970$~kpc) Local Group dwarf galaxy WLM measured from color-magnitude diagrams constructed from deep Hubble Space Telescope imaging. Our observations include a central ($0.5 \, r_h$) and outer field ($0.7 \, r_h$) that reach below the oldest main sequence turnoff. WLM has no early dominant episode of star formation: 20\% of its stellar mass formed by $\sim 12.5$ Gyr ago ($z\sim5$). It also has an SFR that rises to the present with 50\% of the stellar mass within the most recent 5~Gyr ($z<0.7$). There is evidence of a strong age gradient: the mean age of the outer field is 5 Gyr older than the inner field despite being only 0.4~kpc apart. Some models suggest such steep gradients are associated with strong stellar feedback and dark matter core creation. The SFHs of real isolated dwarf galaxies and those from the the Feedback In Realistic Environment suite are in good agreement for $M_{\star}(z=0) \sim 10^7-10^9 M_{\odot}$, but in worse agreement at lower masses ($M_{\star}(z=0) \sim 10^5-10^7 M_{\odot}$). These differences may be explainable by systematics in the models (e.g., reionization model) and/or observations (HST field placement).  We suggest that a coordinated effort to get deep CMDs between HST/JWST (crowded central fields) and WFIRST (wide-area halo coverage) is the optimal path for measuring global SFHs of isolated dwarf galaxies.
\end{abstract}




\section{Introduction}
\subsection{Star Formation Histories of Isolated Dwarf Galaxies}
Isolated dwarf galaxies are key to understanding low-mass galaxy evolution. By virtue of their isolation, a balance between internal processes (e.g., feedback, galactic winds, outflows) and universal external drivers (e.g., gas accretion, cosmic reionization) alone govern their evolution. This picture stands in stark contrast to satellites, whose evolutionary trajectories and present day properties have been strongly affected by their proximity to a massive galaxy, making it challenging to disentangle the relative contributions of environment (e.g., ram pressure, tidal effects) from secular processes. 

Our empirical knowledge of low-mass galaxy evolution has increased markedly over the last decade. Ambitious investments in stellar spectroscopy \citep[e.g.,][]{leaman2013, kirby2017} and imaging programs such as Local Cosmology with Isolated Dwarf program \citep[LCID,][and references therein]{gallart2015}, the ACS Nearby Galaxy Survey Treasury \citep{dalcanton2009}, and archival efforts \citep[e.g.,][]{holtzman2006} have demonstrated a rich diversity in the star formation histories (SFHs) of isolated dwarf galaxies over cosmic time \citep[e.g.,][]{cole2007, hidalgo2009, monelli2010b, monelli2010c, weisz2011a, cole2014, skillman2014, weisz2014a}. These findings reinforce the long-standing theoretical concepts that the evolution of low-mass galaxies are sensitive to a variety of processes including inhomogeneous reionization, variable mass assembly histories, and stellar feedback.

Additionally, large gains in computing power and resolution have led to transformative improvements in numerical models of dwarf galaxy evolution, providing the opportunity to test these refined predictions observationally \citep[e.g.,][and references therein]{elbadry2016, sawala2016, elbadry2017, fitts2017, robles2017, gonzalezsamaniego2017, fillingham2018, fitts2018, su2018, garrisonkimmel2019}.

The large diversity in the evolutionary histories of isolated dwarfs necessitates a larger sample of galaxies with deep color-magnitude diagrams (CMDs) to establish `typical' behavior and its degree of variation. However, the desire to obtain more data is counterbalanced by the observational resources required to reach the faint oldest main sequence turnoff (MSTO) -- the `gold-standard' CMD depth for measuring a well-constrained SFH over cosmic time \citep[e.g.,][]{Gallart:2005qy, dolphin2012}. Due to the high observational demands, only seven isolated dwarf galaxies to date have sufficiently deep observations for well-constrained SFHs, and the sample size has only been incrementally increasing.

Perhaps the most intriguing result from the existing studies is the possibility that isolated dwarf galaxies experience a significant delay in producing their stars relative to satellite galaxies. This was first hinted at from the HST observations of Leo~A \citep{tolstoy1998, cole2007}, which indicated that over 90\% of its star formation occurred in the last 8 Gyr.
This notion was supported by the \citet{cole2014} study of the Aquarius dwarf irregular galaxy (DDO~210), where HST observations have shown that only 10\% of the stars formed earlier than 10 Gyr ago, with a strong increase in the star formation rate between 6 and 8 Gyr ago.
Further tests of this hypothesis are clearly needed.

\subsection{WLM}
\label{sec:wlm}

In this paper, we present the SFH of the isolated, gas-rich, dwarf galaxy Wolf-Lundmark-Melotte (WLM, also known as DDO 221) based on deep HST imaging. Table \ref{tab:wlm_properties} lists the basic properties of WLM. 

At $\sim 970$ kpc, WLM is one of the most distant and isolated dwarf galaxies with such deep HST imaging. HST's spatial extent is sufficient to sample the inner and outer regions of WLM, providing a handle on both the global and radial SFH over cosmic time.
Early resolved-star observations of WLM \citep{ferraro1989} indicated a relatively constant
star formation rate over its lifetime. \citet{minniti1996, minniti1997} demonstrated that 
the oldest stars were more widely distributed than intermediate age and young stars, indicating
a clear radial gradient in the distributions of its stellar populations. SFHs measured from shallow WFPC2 data show that the central regions of WLM are predominantly young \citep{dolphin2000a, weisz2014a}.

Existing metallicity measurements within WLM can serve as consistency checks on the 
age-metallicity relationships that we derive from the stellar photometry.
\citet{leaman2013} presented a detailed study of 180 WLM red giant branch stars and found
a mean value of [Fe/H] $=$ $-$1.28 $\pm$ 0.02 with a dispersion of 0.38 $\pm$ 0.04 dex. This relatively low average metallicity for the older stellar population is typical for
a dwarf galaxy of this stellar mass. The young stellar and ISM metallicities are also
relatively low and typical for galaxies of comparable stellar mass;  \citet{urbaneja2008}
found an average metallicity of $[Z] = -$0.87 $\pm$ 0.06 from 8 young supergiant stars in good agreement with the HII region oxygen abundances measured by \citep{skillman1989, hodge1995, lee2005}.

A few words about the degree of isolation of WLM are appropriate here. WLM is currently slightly more distant from the Milky Way than from M31 \citep[D$_{M31} =$ 836 kpc;][]{mcconnachie2012} and lies near the zero-velocity surface for the Local Group. From a comparison of positions and radial velocities, \citet[][]{teyssier2012} conclude that there is a less than 1\% probability that WLM is associated with the Milky Way. To our knowledge, no similar analysis has been conducted for the WLM-M31 association, but the large current separation indicates a very low probably of past interaction. 

WLM's large distance also places it beyond the so-called ``backsplash'' radius \citep[e.g.,][]{gill2005}. Galaxies beyond this distance, nominally twice the virial radius of the central system (i.e., 600~kpc for the MW), are unlikely to have been ejected from an early interaction with the central galaxy (though see \citealt{donghia2009} which suggests Cetus and Tucana may exist as the results of early interactions). Around the LG, only six galaxies more distant than the backsplash radius have SFHs measured from deep CMDs \citep[e.g.,][]{cole2014, gallart2015}.

This paper is organized as follows. In section \ref{sec:methodology}, we describe the observations, data reduction, distance determination, and method by which we measured the SFH. In section \ref{sec:results} we present the SFH of the inner, and outer fields of WLM. In section \ref{sec:discussion} we compare WLM's SFH with other isolated dwarf galaxies and to simulated dwarf galaxies. In the appendices, we include tabulated SFHs for WLM and demonstrate the weak effect of stellar model choice on the resulting SFH of WLM.

This is the first paper in a series aimed at comparing the global and spatial properties of isolated dwarf galaxies to state-of-the-art models of low-mass galaxy formation. Accordingly, the scope of this first paper is to present the CMD and SFH of WLM, consider its global SFH relative to other isolated galaxies and select simulations, and briefly comment on radial trends, which will be explored more depth in a future paper in this series.

\begin{figure}
	\includegraphics[width=\columnwidth]{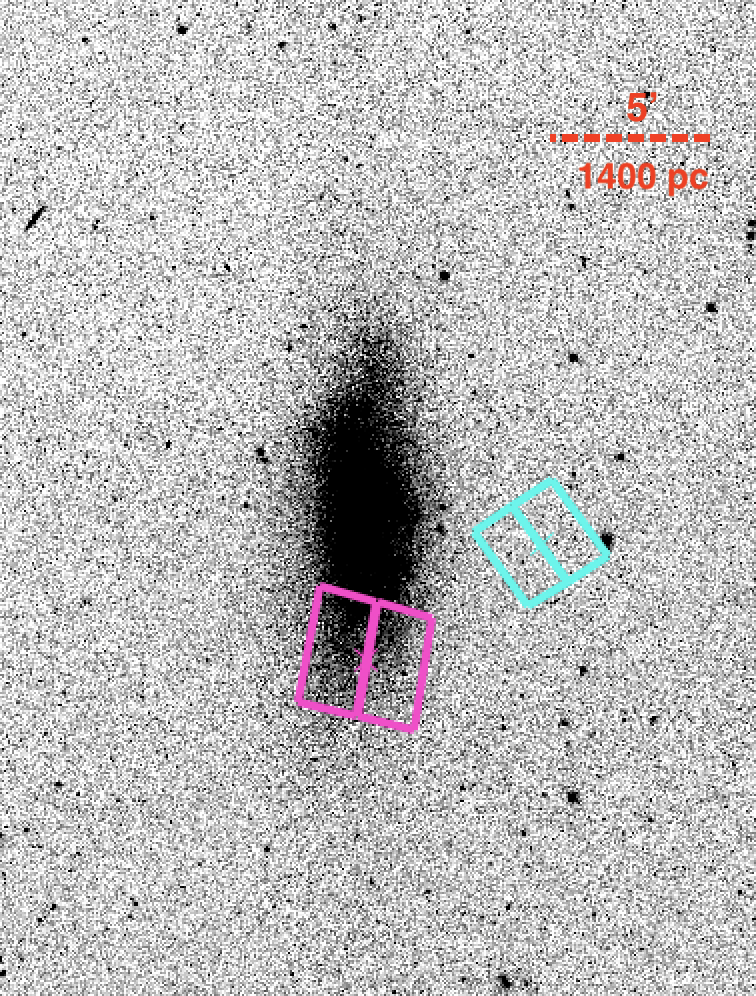}
    \caption{The two HST fields overlaid on a ground-based image of WLM.  The central ACS field is shown in magenta and the outer UVIS field in cyan.  The fields were placed to sample spatial gradients and provide a reasonable representation of WLM's global SFH.}
    \label{fig:spatial_fig}
\end{figure}

\section{Methodology} 
\label{sec:methodology}

\subsection{Observations and Photometry}

The observations and data reduction for WLM mirror that of several previous HST programs aimed at isolated dwarf galaxies. Here, we briefly summarize the data acquisition and reduction, and refer the reader to more detailed descriptions in \citet{monelli2010b}, \citet{cole2014}, and \citet{skillman2014}.

WLM was observed with HST between July 17 and July 19, 2015 as part of  HST-GO-13768 (PI D. Weisz). An inner ACS and outer UVIS field were observed for $\sim$27,360s in F475W and $\sim$ 34,050s in F814W. Figure \ref{fig:spatial_fig} shows the placement of the HST fields overlaid on a ground-based optical image. Two fields were selected to sample the population gradients in WLM. The inner ACS field is located 1~kpc (0.5 $r_h$) from the photometric center, while the outer UVIS field is 1.4~kpc (0.7 $r_h$) from the center. At the distance of WLM, the fields are 0.95~kpc (ACS) and 0.76~kpc (UVIS) across in linear size.

We measured photometry of stars in the ACS and UVIS fields using the point spread function (PSF) fitting package \texttt{DOLPHOT} \citep{dolphin2000b}, with specific ACS and UVIS modules, using the photometric reduction parameters recommended by \citet{williams2014}.

\begin{table}
\begin{center}
\caption{Properties of WLM}
\begin{threeparttable}
\label{tab:wlm_properties}
\begin{tabular}{ lc }
\toprule[1pt]\midrule[0.3pt]
Parameter & Value  \\
\hline
Galactic Coordinates ($\ell$,$b$) [degree] & 75.9, -73.6 \\
Distance (kpc) & $968^{+5, 41}_{-7, 39}$ \\
Distance Modulus (mag) & $24.93^{+0.02, 0.07}_{-0.02, 0.07}$ \\
Absolute Magnitude ($M_V$)& -14.2 $\pm$ 0.01  \\
Line of sight reddening E(B-V)& 0.038 \\
Half-light radius ($^{\prime}$)& 8.62 $\pm$ 0.26 \\
Stellar Mass ($M_\odot$)& 4.3 $\times$ $10^7$   \\
Current SFR (log($M_\odot \ yr^{-1}$)) & -2.24  \\
Stellar Metallicity (dex) &-1.28 $\pm$ 0.02   \\
\hline
\end{tabular}
\begin{tablenotes}
      \small
      \item Note: All values are taken from \citet{mcconnachie2012}  except for mean metallicity, which is taken from \citet{leaman2013}, and the SFR which is from \citet{karachentsev2013}. The distance is from this work.  Uncertainties on the distance reflect the random and systematic errors.
    \end{tablenotes}
  \end{threeparttable}
\end{center}
\end{table}

From the raw source catalogs, we culled the photometry to only include high-fidelity stellar sources. We required that stars have a SNR$>5$ in both filters, $sharp_{F475W}^2 + sharp_{F814W}^2 < 0.1$ and $crowd_{F475W} + crowd_{F814W} < 1.0$.

To quantify observational uncertainties and completeness we ran $\gtrsim 10^5$ artificial star tests (ASTs) in each field.  The 50\% completeness limits for the ACS field are $m_{F475W} = 28.85$, $m_{F814W} = 28.17$ and for the UVIS field are $m_{F475W} = 29.20$, $m_{F814W} = 28.15$.

\subsection{Color-Magnitude Diagrams}
\begin{figure}
	\includegraphics[width=\columnwidth]{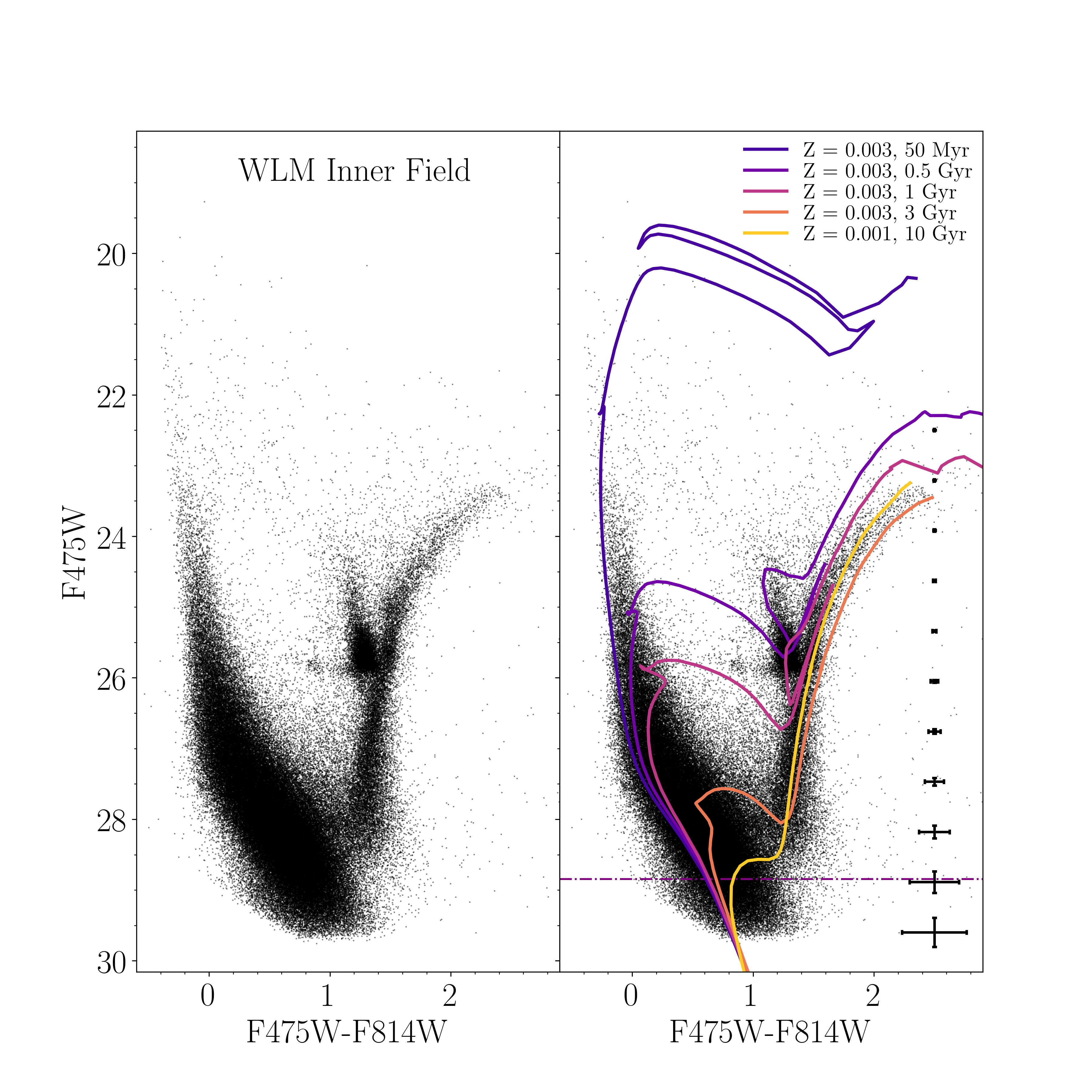}
    \caption{CMD of the inner ACS field of WLM. Virtually all regions of the CMD are well-populated, suggesting quasi-continuous star formation over the lifetime of WLM.  In the right panel, we overplot select Padova isochrones that confirm this scenario.  The 50\% completeness limit (dashed line) in F475W of the ACS field is $\sim 0.5$ mag below the oldest MSTO, allowing for excellent leverage on the SFH at all lookback times.}    
    \label{fig:acs_cmd}
\end{figure}

\begin{figure}
	\includegraphics[width=\columnwidth]{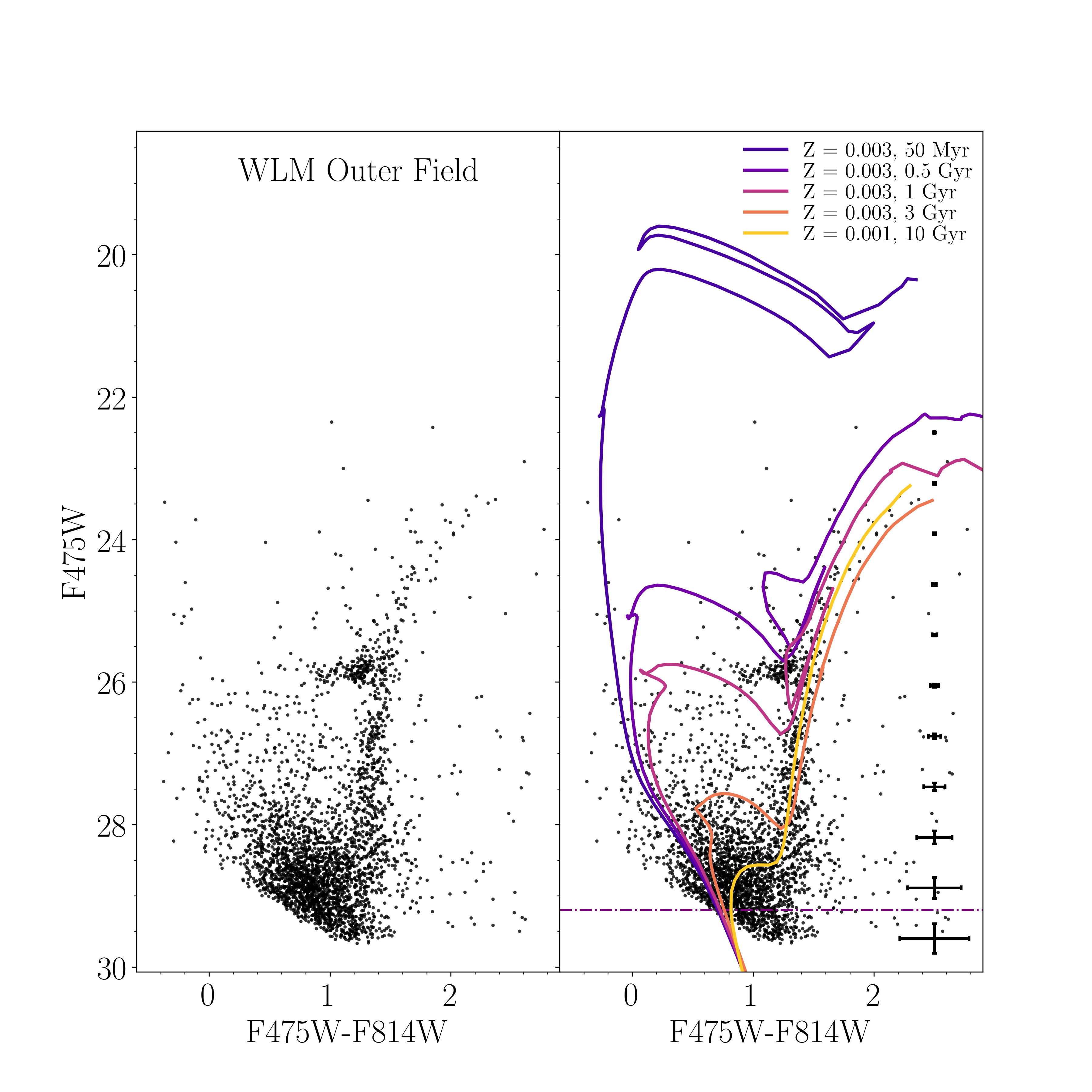}
    \caption{CMD of the outer UVIS field of WLM. Compared to the ACS field, this field is more sparsely populated and lacks stars on the upper main sequence suggestive of little to no recent star formation. As with the ACS field, the F475W 50\% completeness limit is $\sim 0.5$ mag below the oldest MSTO, allowing for excellent leverage on the SFH at all lookback times.}
    \label{fig:uvis_cmd}
\end{figure}

Figures \ref{fig:acs_cmd} and \ref{fig:uvis_cmd} show the CMDs for the ACS and UVIS fields, respectively. The ACS field has 149,558 stars and all stellar sequences are well-populated and clearly visible. Visual inspection shows subgiant stars for every isochrone age and suggests that the inner ACS field has had continuous star formation at virtually all epochs. The presence of stars at oldest MSTO, as well on the horizontal branch (HB) in each field indicate ancient and intermediate age star formation.  The red HB is clearly visible in each CMD, while blue HB merges into the luminous MS, making it challenging to discern visually. The red HB and red clump indicate intermediate age star formation, and the luminous MS and blue and red core helium burning sequences (BHeBs, RHeBs) are signs of star formation within the most recent 1 Gyr. 

We plot select Padova isochrones \citep{girardi2010} in the right-hand panel to guide the reader's eye. For select ages (and metallicities) of 0.05, 0.5, 1, 3, and 10 Gyr, we find stars that populate each area of the isochrone, which is a sign of quasi-continuous star formation at these ages.  It is challenging to visually interpret the relative densities of stars for each isochrone, i.e., which would indicate bursts of star formation, and we therefore defer detailed discussion of putative features until we present a quantitative SFH in \S \ref{sec:acs_sfh}.

The outer UVIS CMD (Figure \ref{fig:uvis_cmd}) contains 3,362 stars and is not as well-populated as the ACS field, by virtue of its location along WLM's minor axis. However, even with fewer stars, many of the same features (e.g., oldest MSTO, HB, red clump) are clearly defined. Compared to the ACS CMD, the UVIS CMD has few, if any, stars younger than $\sim$ 500 Myr.  We overplot select Padova isochrones in the right hand panel of Figure \ref{fig:uvis_cmd}.  For ages $> 3$~Gyr, stars appear to populate all of the selected isochrones, suggesting continuous star formation for ages older than $\sim 3$~Gyr.  For ages younger than $\sim 3$~Gyr, the interpretation is less clear. While stars do populate portions of the $\lesssim 1$~Gyr isochrones, the sparsity of objects suggests very little, or perhaps no, star formation at young ages.  The visual translation of such sparse populations to a quantitative SFH is challenging in sparse regimes and we thus defer discussion of recent star formation to our measurement of the quantitative SFH in \S \ref{sec:uvis_sfh}.

\subsection{Distance Determination}

The well-populated ACS field allows us to measure a tip of the red giant branch (TRGB) distance to WLM. To compute this, we use an implementation of the maximum likelihood technique first described in \citet{makarov2006}. Briefly, this code constructs a model power-law luminosity function, convolves it with photometric errors and completeness from the ASTs, and iterates over a grid of parameter values to find the apparent magnitude of the TRGB.

 We find the magnitude of the TRGB to be $m_{F814W} = 20.91_{-0.01}^{+0.02}$, where the uncertainties reflect the narrowest 68\% confidence intervals that include the most likely value.

We then convert this to a distance modulus by first correcting the apparent magnitude for line of sight Galactic foreground extinction ($A_V = 0.104$) using the maps from \citet{schlafly2011}. We then used the calibration of \citet{rizzi2007}

\begin{equation}
\centering
    M_{F814W}^{ACS} = -4.06 + 0.15 \, [(F555W-F814W)-1.74]
\end{equation}

\noindent to measure the distance modulus.  Because \citet{rizzi2007} do not provide a calibration for $F475W$, we use the color approximation of $(F555W - F814W) = 0.675 \, (F475W - F814W)$. The mean color of the TRGB region in both CMDs is $(F475W-F814W) \sim 2.4$.  

We find that WLM has a TRGB distance modulus of $\mu = 24.93 \pm 0.02 \, ({\rm random}) \pm 0.07 \, ({\rm systematic})$, making WLM the most distant galaxy for which the oldest MSTO has been observed. This places WLM at a distance of $D=968_{-7, 39}^{+5, 41}$~kpc, the first error value representing random error, the second value representing the random and systematic error. This distance agrees with previous distance determinations of 24.93 $\pm$ 0.04 from a TRGB measurement \citep{rizzi2007}, 
24.92 $\pm$ 0.04 $\pm$ 0.04 from Cepheid observations by \citet{gieren2008}, and 24.99 $\pm$ 0.10 using the flux-weighted gravity-luminosity relationship for A and B supergiants \citep{urbaneja2008}. We adopt $\mu=24.93$ for subsequent analysis in this paper.

\subsection{Measuring the Star Formation History}
\label{sec:methods}

To measure the SFH of WLM, we model the CMDs of the ACS and UVIS fields using \texttt{MATCH} \citep{dolphin2002}. \texttt{MATCH} has been widely used to measure the SFHs of $> 100$ galaxies throughout the LG and Local Volume. In this analysis, we generally follow the \texttt{MATCH} usage as described in \citet{weisz2014a}. Here, we briefly summarize how \texttt{MATCH} operates and how we applied it to the WLM CMDs.

For a given set of stellar evolution models, stellar initial mass function, binary fraction, distance modulus, extinction model, and SFH (i.e., star formation rate and metallicity history as a function of time that is simply the sum of simple stellar populations), \texttt{MATCH} constructs a synthetic CMD. The synthetic CMD is convolved with the error and completeness functions determined by the ASTs, and CMDs of foreground and background contaminants are linearly added to the galaxy synthetic CMD. This mock observed CMD is then compared to an observed CMD using a Poisson likelihood function. The process is repeated for various SFHs (i.e., different weights on each of the simple stellar populations) until a global, most likely SFH is determined.

\begin{figure}
	\includegraphics[width=\columnwidth]{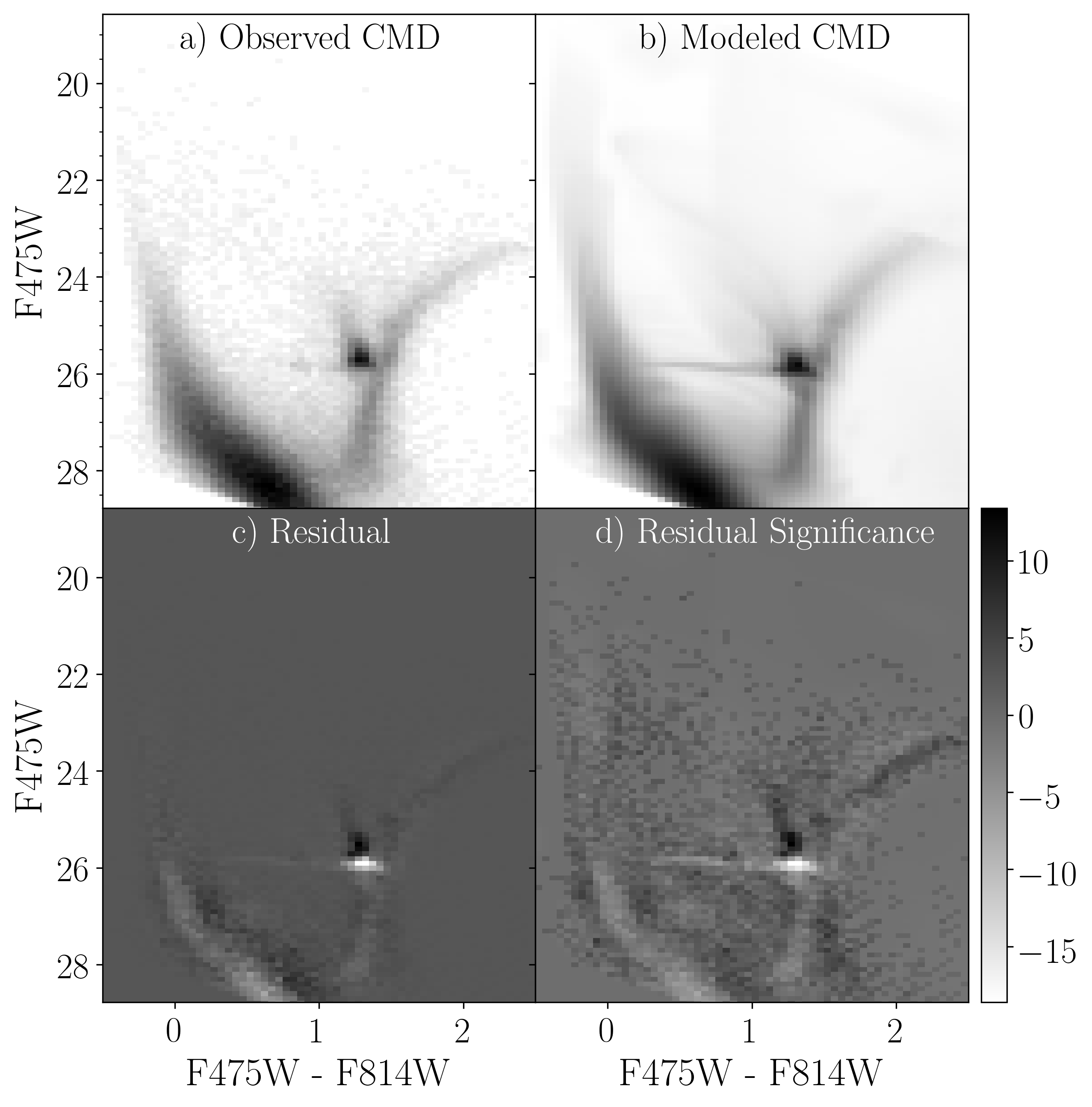}
    \caption{The residual CMDs for our best fit SFH of the ACS field. Panel (a) shows the observed Hess diagram; Panel (b) shows the best fit model CMD; Panel (c) is the residual (i.e., data-model); and Panel (d) is the residual significance, i.e., the residual weighted by the variance in each pixel.  The scale bar is in units of standard deviation. The fit quality is typical of resolved isolated dwarf galaxies. The model is generally good, with known deficiencies in areas of evolved stars (e.g., lower blue loop, RGB, RC, and HB). The bulk of the residual mismatches are due to a spatially varying noise model as discussed in \S \ref{sec:methods}}
    \label{fig:acs_residual}
\end{figure}

For our analysis of WLM, we used the Padova \citep{girardi2010} stellar evolution models, a \citet{kroupa2001} IMF, a binary fraction of 0.35 with the primary-to-secondary mass ratios drawn from a uniform distribution, our TRGB distance of $\mu = 24.93$, and a foreground extinction value of $A_V =0.104$ from \citet{schlafly2011}. We use an age grid of $\log(t) =$ 6.6 to 10.15 with a time resolution of 0.1 dex for $log(t) \le 9.0$ and 0.05 dex for $log(t) > 9.0$. Our metallicity grid is [M/H] = -2.3 to -0.1 with a resolution of 0.1 dex. We model the entire CMD (i.e., do not exclude any regions such as the RGB).

Though more recent stellar models than Padova are now available \cite[e.g., PARSEC, MIST, BaSTI;][]{bressan2012, choi2016, hidalgo2018}, we adopt the Padova models for direct comparison to other isolated galaxies, which all have SFHs measured with these older models. We compare the results of fitting different models to the CMD in Appendix \ref{sec:appendb}. Future papers in this series will measure SFHs for all isolated galaxies using updated stellar models.

\begin{figure}
	\includegraphics[width=\columnwidth]{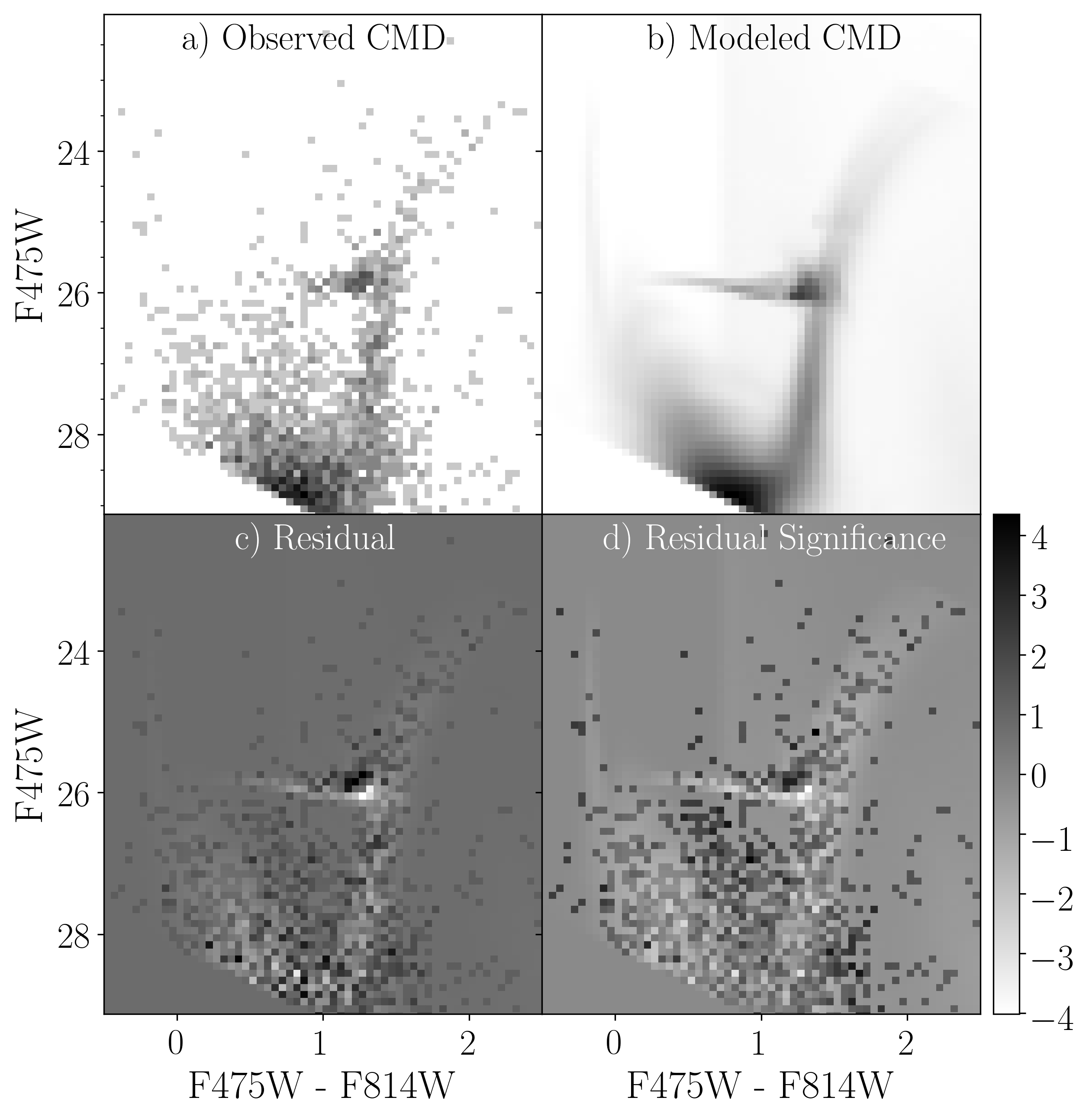}
    \caption{Same as Figure \ref{fig:acs_residual}, except for the UVIS field. The residual appears cleaner than the ACS field due to the simpler nature of the stellar population and having a noise model that doesn't strongly vary across the field due to no surface brightness gradient.}
    \label{fig:uvis_residual}
\end{figure}

We also consider the effects of differential extinction on the SFH. Specifically, beyond the foreground dust value, \texttt{MATCH} reddens a uniform fraction of all model stars between 0 and 0.5 mag by steps of 0.1. We computed the SFH with this differential extinction model in both the ACS and UVIS field, but found that both fields were best fit with no differential extinction.  

While differential extinction may not affect the majority of the CMD, it can affect only the younger stars \citep[e.g.,][]{dolphin2003} which may reside in dusty star formation regions. Following \citet{dolphin2003}, we adopt an age-dependent differential extinction model for young stars such that stars younger than 40~Myr randomly get a value of $A_V=0.5$ applied. The amount of dust applied to the model CMD decreases linearly from $A_V=0.5$ to $A_V=0.0$ between 40 and 100~Myr. Not including age-dependent young star dust results in a model upper main sequence that is too narrow compared to the data.

Using the best fit SFH as a starting point, we compute random and systematic uncertainties on the SFH. Random uncertainties capture plausible variations from the best fit SFH due to finite sampling on the CMD, and are determined using a Hybrid Markov chain Monte Carlo (HMC) algorithm as described in \citet{duane1987} and implemented by \citet{dolphin2013}. Throughout this paper, the reported random uncertainties represent the narrowest 68\% confidence interval around the best fit SFH.

\begin{figure}
\centering
	\includegraphics[scale=0.4]{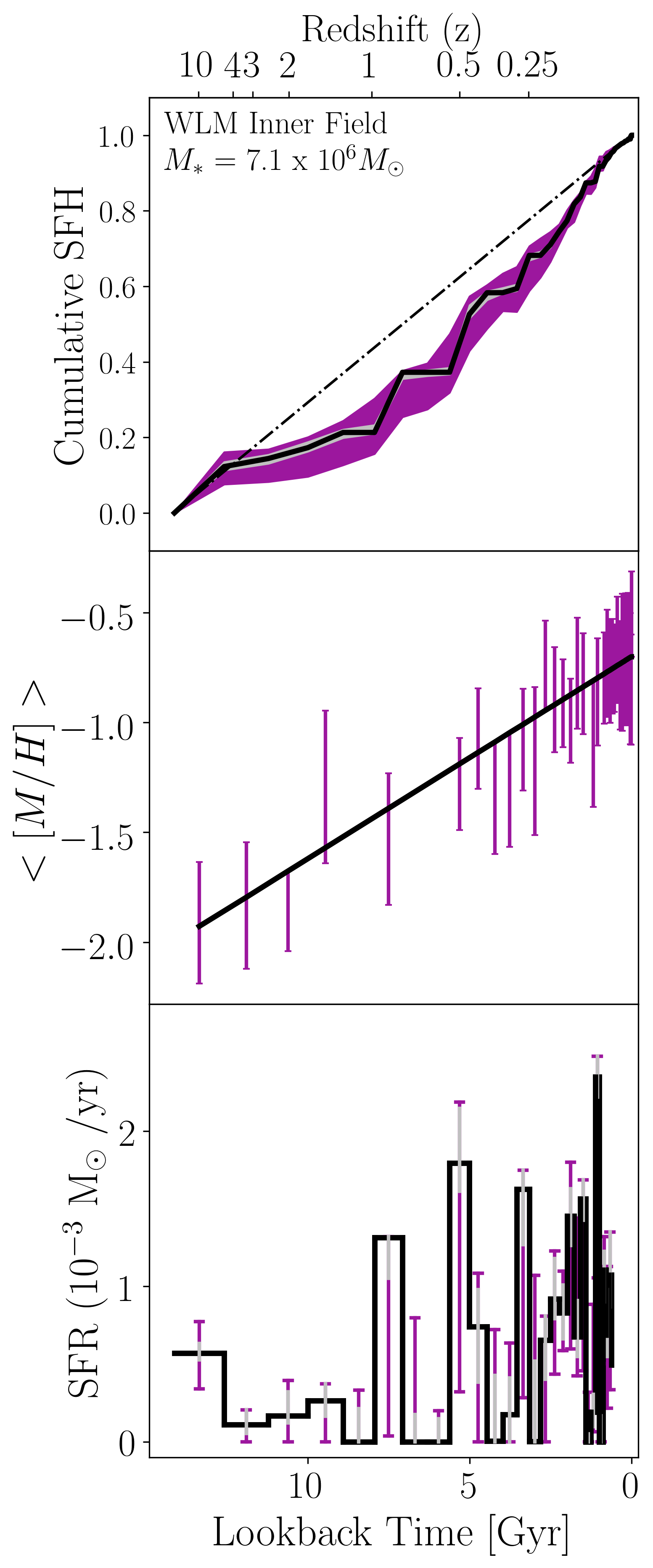}
    \caption{The SFH of the ACS field.  \textbf{Top:} Cumulative Star Formation History of Inner (ACS) Field vs. Lookback Time. The black line reflects the best fit, the grey envelope reflects random uncertainties, and the purple envelope represents total uncertainties (systematic plus random). \textbf{Middle:} Mean Metallicity (log(M/H)) vs. lookback time with random uncertainties in grey and total uncertainties (systematic plus random) in purple, as measured from fitting the CMD.  Only metallicity points in which the best fit SFR is greater than zero are plotted. \textbf{Bottom:} SFR vs. lookback time. About 20\% of the stars in the inner field formed prior to $\sim$12 Gyr ago, and 50\% formed prior to $\sim$ 5 Gyr ago. As in the panels above, the random errors are in grey and the total errors (systematic plus random) are in purple.}
    \label{fig:acs_sfh}
\end{figure}

Systematic uncertainties are designed to approximate variations in the SFH due to the choice of underlying stellar evolution model. We compute systematic uncertainties using 50 Monte Carlo realizations following the procedures described in \citet{weisz2011a}, \citet{dolphin2012}, and \citet{weisz2014a}. The reported systematic uncertainties represent the narrowest 68\% confidence interval around the best fit SFH.

Figures \ref{fig:acs_residual} and \ref{fig:uvis_residual} show the quality of our best fit CMDs for the ACS and UVIS fields, respectively. In both figures, the most important diagnostic is the residual significance (Panel (d)), which is the residual (data-model) weighted by the variance in each pixel. A checkerboard pattern of grey indicates no major residuals, while areas of white or black indicate over/under predictions by the model. The residuals for the ACS and UVIS fields are typical for these types of observations. That is, the model for the ACS field, with more stars and a more complex SFH, is a good match with clear deficiencies. Notably, the areas around the HB and RC are not well-modeled. In part, this is due to imposing a fixed distance. The SFH code would prefer a distance that maximizes agreement with the RC (i.e., it matches more data points). This would place the TRGB well-above the observed TRGB location, thus violating a gold-standard distance measurement. Other areas of disagreement, such as on the RGB, are typical of the Padova models which tend to produce overly blue models (too hot) relative to the data \citep[e.g.,][]{Gallart:2005qy}. Finally, the level of disagreement on the lower part of the luminous MS may be linked to poor matches of the RHeBs, which are known to be problematic for stellar models \citep[e.g.,][]{mcquinn2011}.

\begin{figure}
\centering
	\includegraphics[scale=0.3]{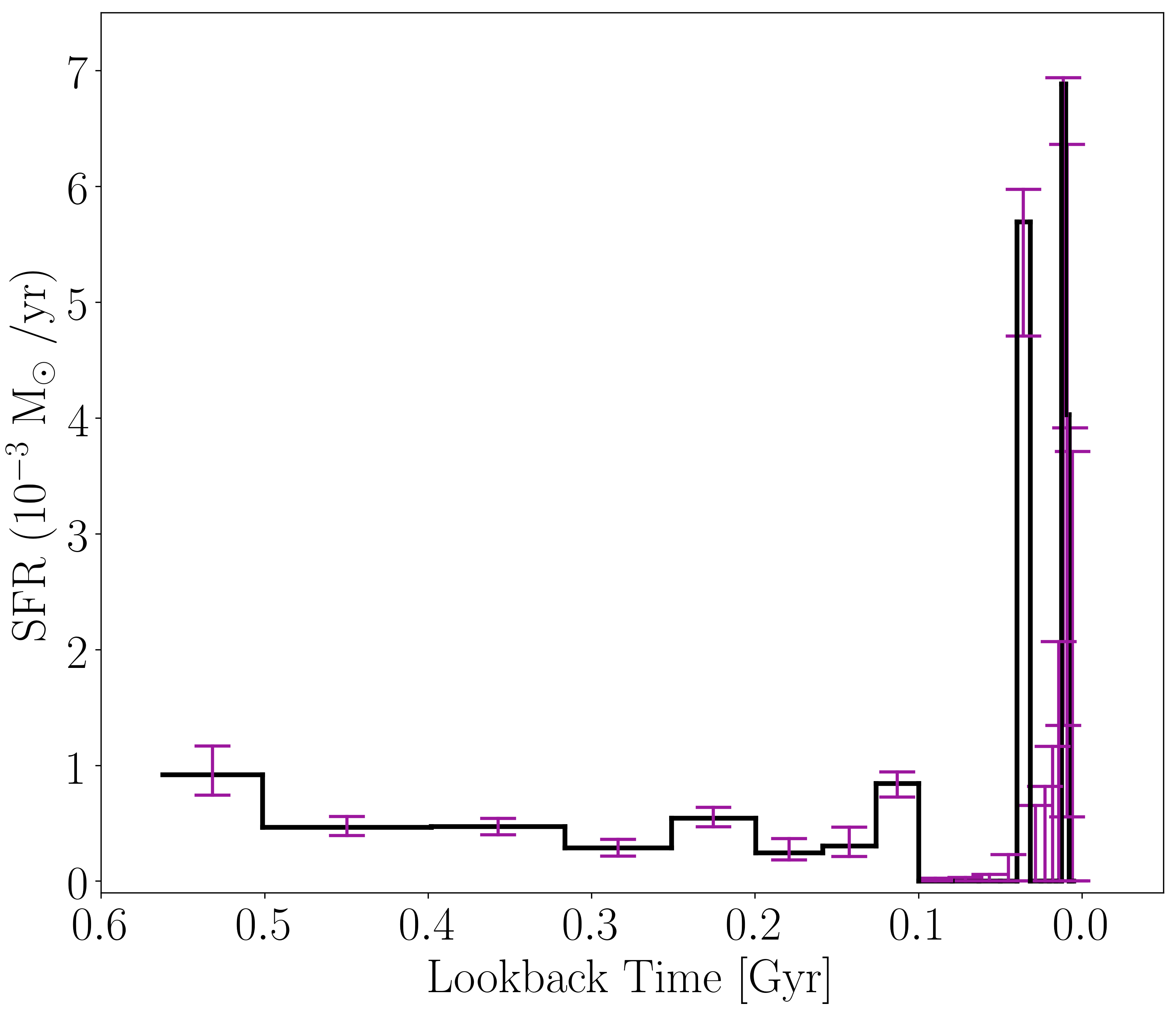}
    \caption{The absolute SFH of the ACS field over the most recent 500 Myr.  In general, the SFH over the last 500 Myr has been fairly constant, with larger amplitude bursts within the more recent 50 Myr.}
    \label{fig:acs_recent_sfh}
\end{figure}

Beyond issues with stellar models, spatial variations in noise model, as determined from ASTs, can affect the quality of the fit. Given that the ACS field covers a large dynamic range in surface brightness (see Figure \ref{fig:spatial_fig}), some of the residuals may be due to mismatch in the noise model.

To test this, we divided the ACS field in a $3\times 3$ grid, and measured the SFH in each of the nine regions using photometry and ASTs limited to each region. For this exercise, we fixed the distance and foreground extinction to common values, but allowed the differential extinction to vary.

Visual inspection of residual CMDs for each of the nine regions reveal notable improvement in the residuals relative to one fit to the entire ACS field. Many of the large scale systematic features that are apparent in the fit to the entire ACS field are absent in the individual regions. Fits to individual regions were consistent with no differential extinction, same as the SFH for the entire field.

Finally, we computed a total SFH from the nine individual regions by first running random and systematic uncertainties on each region and then by combining the nine best fit SFHs and their associated uncertainties. The resultant SFH from this exercise is within a few percent (i.e., consistent within random uncertainties) of the SFH from analyzing the entire field at once.

\begin{figure}
\centering
	\includegraphics[scale=0.4]{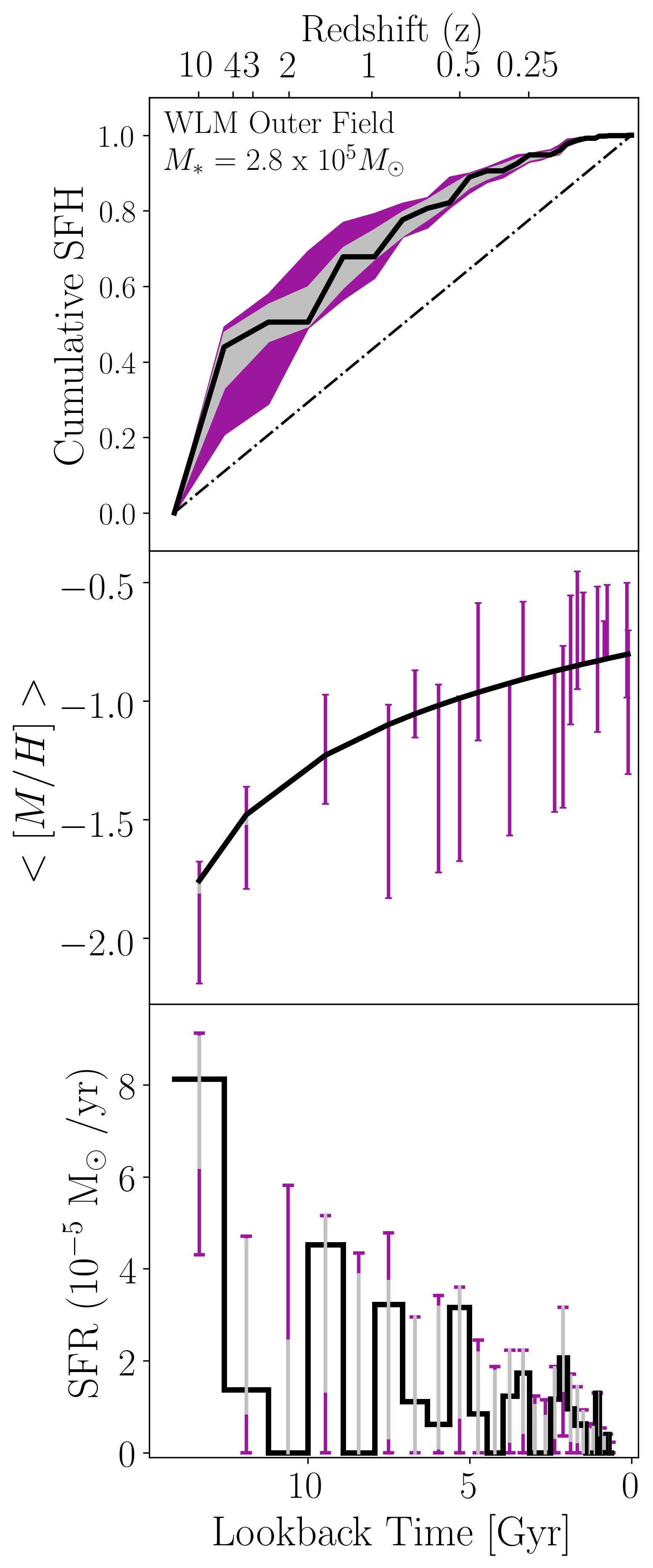}
    \caption{The same as Figure \ref{fig:acs_sfh} only for the SFH of the outer UVIS field. In comparison to the inner field, this outer field formed 50\% of its stars by $\sim$12 Gyr ago and has had little star formation in the last few Gyrs.}
    \label{fig:uvis_sfh}
\end{figure}

This exercise illustrates two points. First, the bulk of the poor residual areas in Figure \ref{fig:acs_residual} are due to a spatial gradients in the noise model rather than issues with the physical model. Second, the consistency of the SFHs indicates that the entire field solution is robust.

Our model of the UVIS field, Figure \ref{fig:uvis_residual}, is well-matched to the data. There are no significant systematics in the residual CMDs. In part, the simpler stellar populations of this field are easier to model and suffer from less age-metallicity degeneracy than the ACS field. The residual for the UVIS field is better than the ACS field owing to a combination of a simpler SFH and lower variability in spatially dependent completeness.

As has been discussed extensively in the literature, the choice of underlying stellar model can affect the measurement of a SFH from a CMD, though variations are smallest for CMDs that reach below the oldest MSTO \citep[e.g.,][]{Gallart:2005qy, hidalgo2009, weisz2011a, dolphin2012, weisz2014a, skillman2017}. To follow up on this point, we measure the SFHs using multiple stellar libraries. As described in Appendix \ref{sec:appendb}, we find that while SFHs measured with different models vary in excess of random uncertainties, our reported systematic uncertainties accurately encompass this variation.

Figure \ref{fig:acs_recent_sfh} shows the absolute SFH zoomed in on the more recent 500 Myr.  Here, we see that the SFH in the ACS field is nearly constant over $\sim$100~Myr intervals from 100-500 Myr ago, with most fluctuations of order unity.  There is a notable increase in SFR within the last $\sim 50$~Myr, over time intervals of $\sim 10$~Myr.  Fluctuations in the SFR over these short intervals are factors of $\sim 4-6$.

\begin{figure}
\centering
	\includegraphics[scale=0.3]{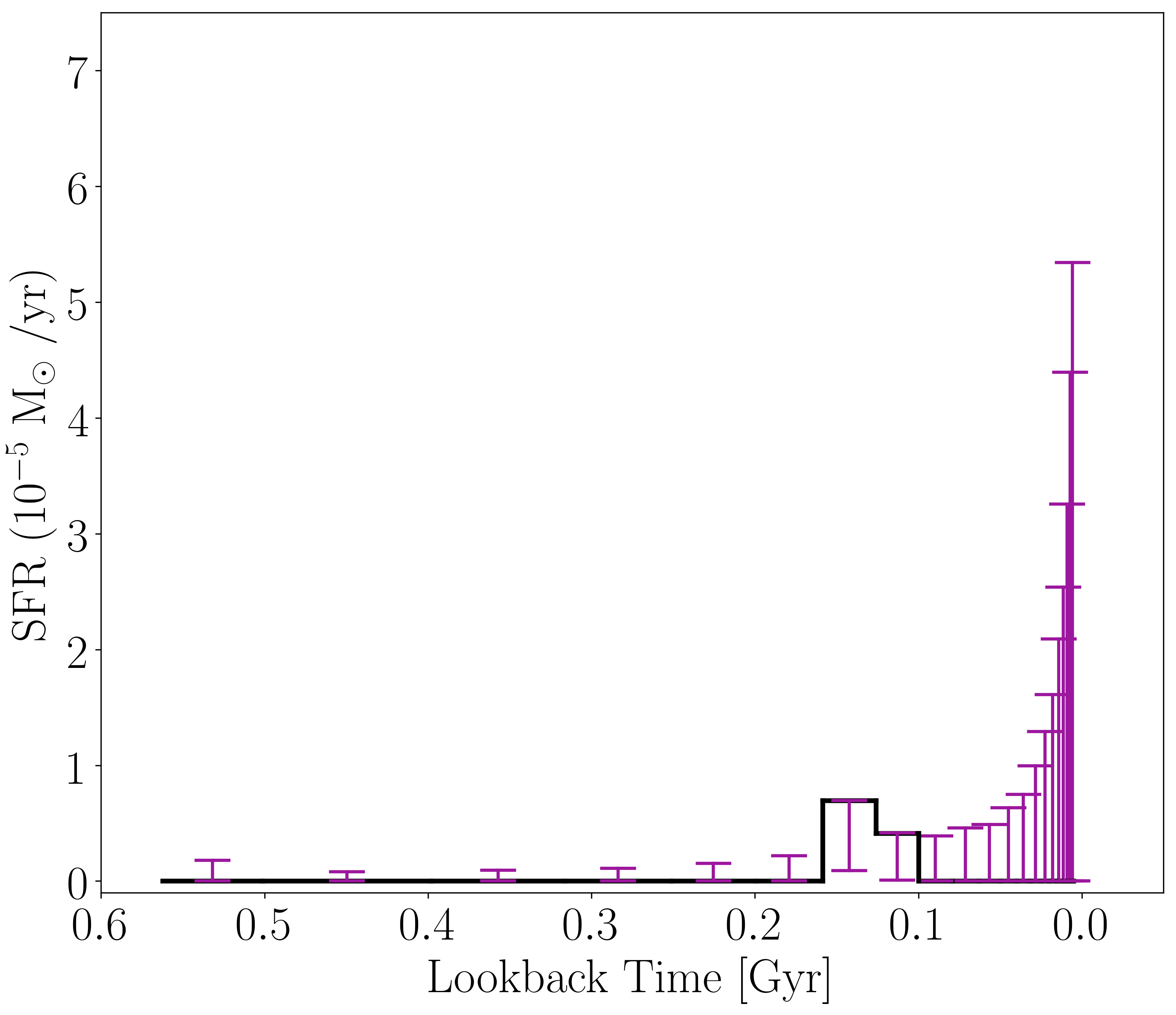}
    \caption{The absolute SFH of the UVIS field over the most recent 500 Myr.  The SFH of this field is consistent with zero over the entire 500 Myr period.}
    \label{fig:uvis_recent_sfh}
\end{figure}

\section{Results}
\label{sec:results}

The main results of this paper are the SFHs of the inner and outer fields in WLM. In the following section, we describe our findings, and in \S \ref{sec:discussion} we place our results into a broader context.

\subsection{Star Formation History of the Inner Field}
\label{sec:acs_sfh}

We first consider the SFH of the inner, ACS field, which is shown in Figure \ref{fig:acs_sfh}.  The top panel presents the cumulative SFH, i.e, the fraction of stellar mass formed prior to a given time. The solid black line is the best fit SFH, the black errorbars reflect random uncertainties, and the purple errorbars represent the total uncertainties, i.e., random and systematic. From the cumulative SFH, we see that WLM had an initial burst of star formation, followed by lower level star formation that increases in activity in the last several Gyr. More quantitatively, $\sim 20$\% of the stellar mass in this field formed prior to $\sim 12.5$~Gyr ago and 50\% formed prior to $\sim 5$~Gyr ago. The fact that 50\% of the stellar mass formed within the most recent $\sim 5$~Gyr indicates that this is a fairly young region in WLM.

The middle panel of Figure \ref{fig:acs_sfh} shows the metallicity evolution of the ACS field.  Our best fit metallicity exhibits linear enrichment in logarithmic metallicity, i.e., log(Z).   The early populations of WLM were quite metal-poor, $[M/H] \sim -2$. By the time 50\% of the stellar mass formed, the field enriched to $[M/H] \sim -1.2$. Our derived present day metallicity for this field in WLM, $[M/H] \sim -0.8$, is in excellent agreement with the spectroscopic metallicities of the young stars and HII regions (see \S \ref{sec:wlm}). Overall, our derived age-metallicity relationship is in great agreement with that derived from spectra of red giant branch stars by \citet{leaman2013}.

The bottom panel of Figure \ref{fig:acs_sfh} shows the absolute SFR versus lookback time. The patterns of star formation in the absolute SFR match the trends inferred from the cumulative SFH. Importantly, this plot provides a sense of how bursty the star formation in WLM has been. For most of its lifetime, fluctuations in the SFR rarely exceed factors of a few to several. Though the time resolution is only sufficient to probe short timescale bursts in the most recent few hundred Myr, the lifetime SFH rules out sustained order-of-magnitude bursts. The strong bursts appear around $\sim$5-7 Gyr ago. While there is clearly increased activity during this time, the large amplitude of the uncertainties make a detailed deconstruction challenging. Additionally, the SFH of WLM shows no evidence of large gaps, i.e., periods of no star formation. One challenge with interpreting these bursts and gaps is that their relative timing and amplitude are coupled to the chosen stellar model.  For example, in Appendix \ref{sec:appendb}, not all models show a burst of star formation from 5-7 Gyr ago, even though they show a similar overall shape.  A detailed analysis of burstiness, and its dependence on the choice of stellar models requires a level of analysis that is beyond the scope of the present paper.  We plan such a pursuit in a follow up paper, in which we can explore trade offs between real bursts of star formation versus features generated by artifacts in the stellar models.

\subsection{Star Formation History of the Outer Field}
\label{sec:uvis_sfh}

Figure \ref{fig:uvis_sfh} shows the SFH of the UVIS field which is located along the minor axis of the galaxy. The SFH of this field is predominantly old, with the bulk of star formation at earlier times: $\sim$40\% of the stellar mass formed prior to 12.5 Gyr ago, 50\% by 10 Gyr ago, and 90\% by 7~Gyr ago. Overall, star formation in the UVIS field appears to decrease toward the present and resembles a declining $\tau$-model SFH.

The chemical enrichment of this field follows a similar trend as the ACS field. The oldest populations are quite metal-poor with $[M/H] \sim -1.7$, which is consistent with the ACS fields when factoring in uncertainties. By $\sim 5$~Gyr ago, this field enriched to its present day metallicity, within errors, of $[M/H] \sim -0.8$.

The absolute SFH of the UVIS field is plotted in the bottom panel of Figure \ref{fig:uvis_sfh}. Note that the absolute scale is a factor of $\sim 100$ less than the ACS field. While there appear to be a number of bursts and gaps, the small number of stars results in large error bars, making them not statistically significant. 

Figure \ref{fig:uvis_recent_sfh} shows the absolute SFH of the UVIS field plotted over the most recent 500 Myr.  At all times shown, the SFH is consistent with zero, though with large error bars due to Poisson sampling statistics that are captured by the HMC runs.

\section{Discussion}
\label{sec:discussion}

\subsection{Comparison with Star Formation Histories of Other Isolated Dwarf Galaxies}
\label{sec:comp_lcid}

Over the past decade, there has been a concerted effort to measure SFHs of dwarf galaxies beyond the virial radii of the MW and M31. The Local Cosmology with Isolated Dwarfs (LCID) program has been at the forefront of this effort, measuring SFHs for Leo~A, Phoenix, LGS~3, Cetus, Tucana, Phoenix, and IC~1613 (see \citealt{gallart2015} and references therein) from HST-based CMDs that extend below the oldest MSTO. A related program measured the SFH of DDO~210 \citep{cole2014} from comparably deep HST imaging.

\begin{figure}
\centering
	\includegraphics[scale=0.25]{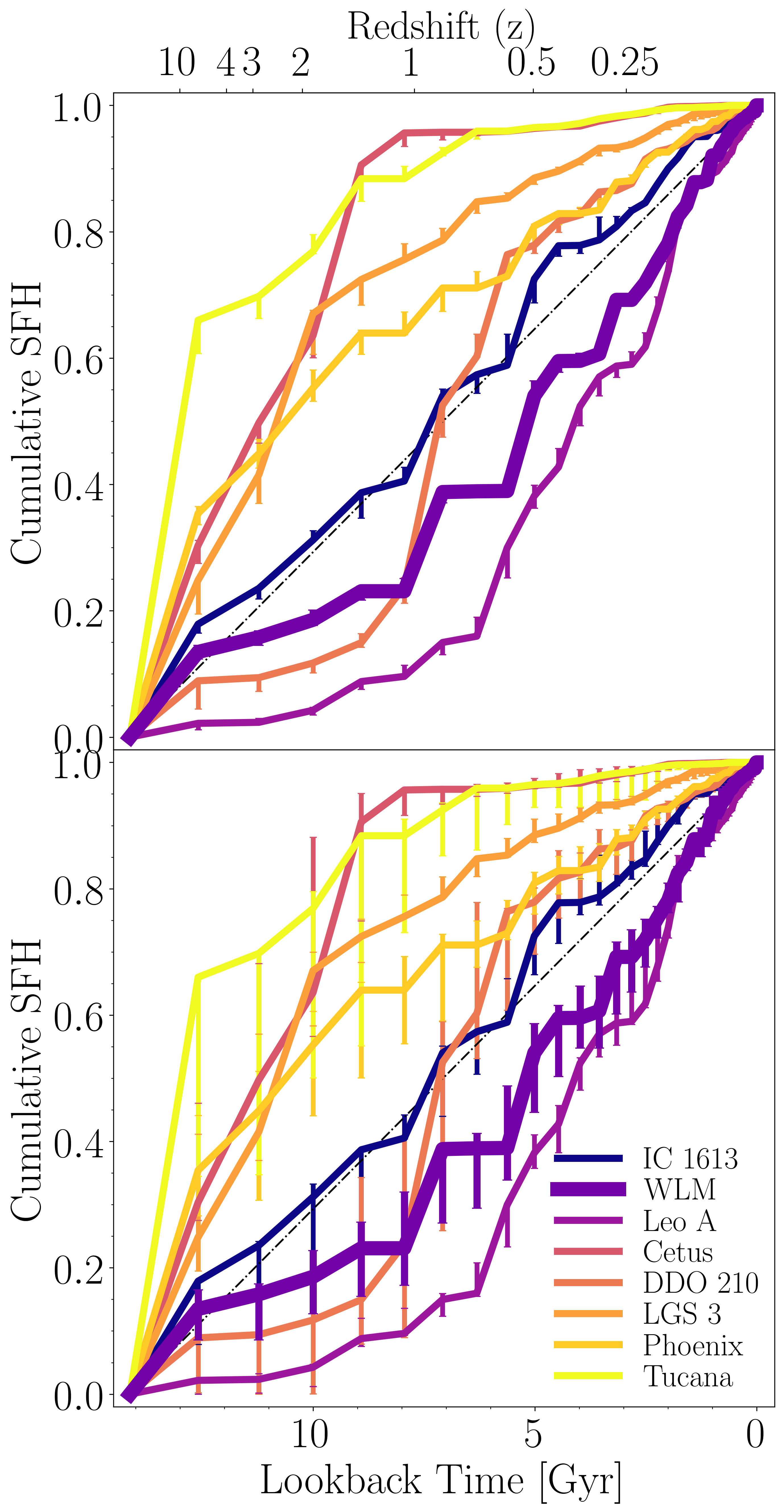}
    \caption{WLM in context. The cumulative SFHs of isolated LG dwarf galaxies, all measured with the Padova models. The upper panel shows best fit SFHs of each galaxy with random errors, while the lower panel includes random and systematic. Galaxy SFHs are color-coded by mass from highest-mass (IC~1613) in navy to lowest-mass (Tucana) in yellow. WLM, the thickest curve in purple, appears to have a rising SFH toward the present.}
    \label{fig:isolated_sfhs}
\end{figure}

A main takeaway from LCID is that, even when removed from the effects of environment, dwarf galaxies exhibit a huge diversity in SFH. This primary finding is illustrated in Figure \ref{fig:isolated_sfhs}, where we plot the cumulative SFHs of the LCID galaxies, DDO~210, and add the ACS SFH of WLM from this paper\footnote{We use the ACS field as a proxy for WLM's SFH because it forms 100 times more stellar mass than the UVIS field, meaning the UVIS field contribution to a ``global'' SFH is negligible relative to ACS}. The lines are colored by the present-day stellar mass of each systems from lowest mass (Tucana, yellow) to highest mass (IC~1613, purple).  

The SFHs in Figure \ref{fig:isolated_sfhs} have all been measured with the same Padova stellar models and the same CMD fitting code \citep[][]{dolphin2002}. In the top panel, we plot the SFHs with only the random uncertainties from the CMD fits, which allows us to gauge the \emph{relative} differences in the SFHs. Because the CMDs are so well-populated, the random uncertainties are small and the SFHs of any two galaxies are inconsistent at a high-level of statistical significance, as discussed in the series of LCID papers. Here, we focus on new information added by WLM.

WLM is the second most massive galaxy in the sample, following IC~1613, making the two interesting to compare. Compared to IC~1613, WLM formed fewer stars at early times.  WLM formed 50\% of its stellar mass by $\sim$5~Gyr ago, whereas IC~1613 formed 50\% of its stellar mass by $\sim$7.5~Gyr ago.  At later times, the fraction of stars formed in WLM is rising with time, while that of IC~1613 is relatively constant.  Thus, WLM formed 90\% of its stellar mass by $\sim$1~Gyr ago, whereas IC~1613 formed 90\% of its stellar mass $\sim$2~Gyr ago.  Note that this comparison is limited by the placement and field size of HST.  The FOV only covers $\sim$10-20\% of the half-light radius in each of the galaxies, and the fields in WLM are all located at $r < 1 r_h$, whereas in IC~1613, the field is located at $r \gtrsim 1 r_h$.  As discussed below, population gradients in both galaxies can affect this type of direct comparison.

Beyond from IC~1613, WLM's SFH is not like any of the other isolated systems. The three lowest-mass systems (Tucana, Phoenix, Cetus) formed the majority of their stellar mass prior to 10 Gyr ago and have had declining SFHs since then. DDO~210, Cetus, and Leo~A are the next three most massive systems, yet their SFHs span nearly the entire range of possibility: Leo-A is mostly young, Cetus is mostly old and DDO-210 is dominated by intermediate-age populations. In short, none are like WLM.

\citet{gallart2015} suggest that SFHs can be separated into two classes of systems, `fast' and `slow', which are a manifestation of halo assembly bias \citep[e.g.,][]{gao2005}. In short, fast systems formed most of their stellar mass at early times because they were located in denser environments in the early Universe. In contrast, `slow' systems were located in less dense environments and formed most of their stellar mass over longer periods of time. In this picture, WLM would be considered a `slow' dwarf as it did not have an early dominant episode of star formation, suggesting it formed in a lower density region like other slow dwarfs such as IC~1613, Leo~A, and DDO~210.  

However, there remain several observational challenges to this interpretation. First is the small sample size. Second, we lack full phase space information. This could be used to determine orbital histories, an important quantity in reconstructing the infall time into the LG and the density of environment at early times. Third, the location of the HST fields vary from galaxy-to-galaxy. Placement of the HST field inside or outside the half-light radius could bias the median age of a system by $\sim 1$ Gyr relative to a true global SFH \citep{graus2019}.

As a useful summary of the isolated dwarf galaxy SFHs, in Table \ref{tab:sfh_summary}, we list the lookback times when 50\% ($\tau_{50}$) and 90\% ($\tau_{90}$) of the stellar mass formed in each galaxy, along with the random and total uncertainties in those values.

\subsection{Comparison to Simulated Dwarf Galaxies}
\label{sec:comp_fire}

There are numerous theoretical models of isolated dwarf galaxy formation and evolution that predict SFHs as a function of various properties such as present day stellar mass \citep[e.g.,][]{governato2010, sawala2010, hopkins2014, shen2014, vogelsberger2014, christensen2016, elbadry2016, sawala2016, fitts2017, wheeler2018, buck2019, wright2019}. While a detailed comparison of measured SFHs to all such models is of great interest, it is a large undertaking that is beyond the scope of this paper. Instead, we select a single model to illustrate one type of comparison that can be made.

\begin{figure}
	\includegraphics[width=\columnwidth]{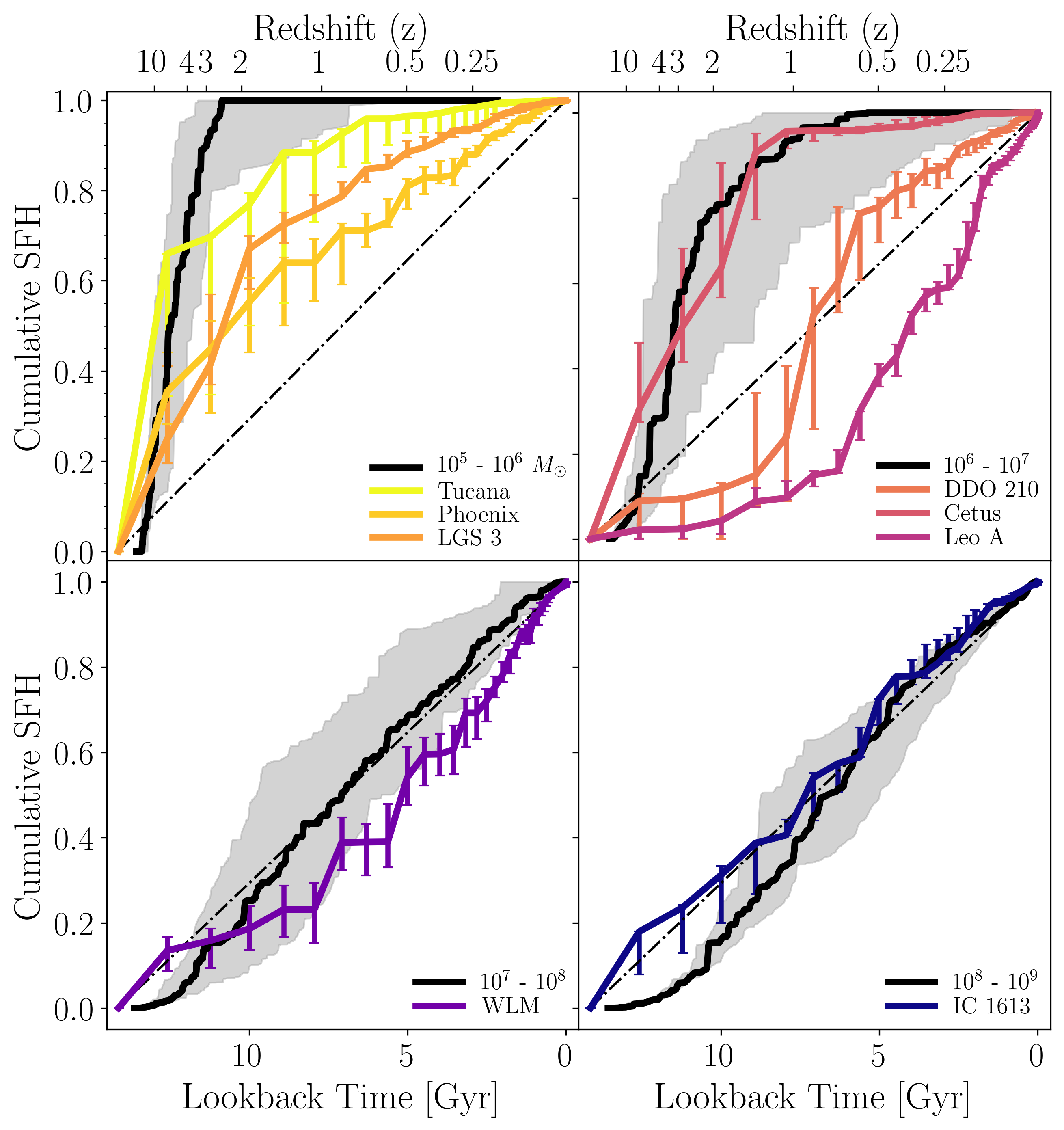}
    \caption{A comparison between the simulated dwarf galaxies from the FIRE suite \citep{garrisonkimmel2019} and real SFHs measured from HST CMDs as a function of present day stellar mass. The solid black lines in each panel are the median of the simulations and the grey bands reflect the 68\% scatter. Uncertainties on the real galaxies include random and systematic components. The comparison is mixed with good general agreement at higher masses and worse agreement at lower masses.}
    \label{fig:fire_sfh_comp}
\end{figure}

Figure \ref{fig:fire_sfh_comp} shows the cumulative SFHs of the real isolated dwarf galaxies from Figure \ref{fig:isolated_sfhs} and median SFHs from the 500 simulated dwarf galaxies analyzed by \citet{garrisonkimmel2019} as part the Feedback in Realistic Environments \citep[FIRE;][]{hopkins2014, hopkins2018} simulation suite.  For the SFHs of real galaxies we plot the systematic uncertainties, i.e., those in the bottom panel of Figure \ref{fig:isolated_sfhs}.

\begin{table}
\caption{Summary Statistics for Isolated Dwarf Galaxy SFHs}
\label{tab:sfh_summary}
\begin{center}
\begin{threeparttable}
\begin{tabular}{l|c|c}
\toprule[1.0pt]\midrule[0.3pt]
Galaxy & $\tau_{50}$ & $\tau_{90}$   \\
\hline
 & (log lookback age) & (log lookback age) \\
\hline
IC 1613 &  9.87$_{-0.00, 0.06}^{+0.00, 0.00}$ &  9.30$_{-0.00, 0.02}^{+0.00, 0.09}$ \\
WLM ACS & 9.71$_{-0.00,0.07}^{+0.01,0.03 }$ & 9.02$_{-0.00, 0.01}^{+0.00, 0.06}$ \\
WLM UVIS & 10.01$_{-0.01,0.02 }^{+0.08, 0.09}$ & 9.67$_{-0.07, 0.10}^{+0.02, 0.03}$ \\
Leo A & 9.61$_{-0.02, 0.03}^{+0.01, 0.01}$ & 8.93$_{-0.01, 0.02}^{+0.02, 0.06}$ \\
Cetus & 10.05$_{-0.01, 0.03}^{+0.00, 0.04}$ & 9.95$_{-0.02, 0.05}^{+0.00, 0.04}$ \\
DDO 210 & 9.85$_{-0.01, 0.02}^{+0.00, 0.03}$ & 9.33$_{-0.01,0.02}^{+0.00, 0.09}$ \\
LGS 3 & 10.03$_{-0.01, 0.01}^{+0.00, 0.04}$ & 9.64$_{-0.01, 0.01}^{+0.02, 0.08}$ \\
Phoenix & 10.03$_{-0.01, 0.08}^{+0.01, 0.03}$ & 9.42$_{-0.02, 0.02}^{+0.01, 0.03}$ \\
Tucana &  10.11$_{-0.00, 0.11}^{+0.00, 0.00}$ & 9.88$_{-0.00, 0.13}^{+0.03, 0.04}$ \\
\hline
\end{tabular}
\begin{tablenotes}
      \small
      \item $\tau_{50}$ is the lookback time when 50\% of the stellar mass formed and $\tau_{90}$ is the lookback time when 90\% of the stellar mass formed. The uncertainties reflect the random and total (random plus systematic) components.
    \end{tablenotes}
  \end{threeparttable}
\end{center}
\end{table}

Following Figure 4 in \citet{garrisonkimmel2019}, we plot the median (black) and 68\% scatter (grey envelope) in the simulated SFHs. We have plotted the SFHs of `centrals in Local Groups,' which are defined as dwarf galaxies between 0.3 and 2~Mpc from the central two galaxies (e.g., MW and M31 analogs) in FIRE realizations of LG environments. As in \citet{garrisonkimmel2019}, we group the real and simulated galaxies by stellar mass.

For the specific case of WLM (lower left panel of Figure \ref{fig:fire_sfh_comp}), we find reasonable agreement between its SFH and predictions from the FIRE simulations. That is, WLM is statistically consistent (when systematics are included) with FIRE simulations in this mass range. The most notable difference is at the earliest epoch, when WLM formed 10-15\% of its stellar mass by 12.5 Gyr ago ($z\sim5$) and the simulations formed 2-6\%. A similar discrepancy is observed at higher masses. By 12.5 Gyr ago, IC~1613 formed 15-20\% of its stellar mass and the simulations formed only 1-2\% of its stellar mass. There are a number of selection effects that could explain this difference (e.g., field placement, small number statistics). However, as discussed in \citet{weisz2014a}, a similar under-prediction in stellar mass formation is  observed in some empirical models that include low-mass galaxies \citep[e.g.,][]{behroozi2013}. While a detailed exploration of the earliest epochs of star formation in real and simulated dwarf galaxies is beyond the scope of this paper, understanding such discrepancies is important because ancestors of galaxies like WLM and IC~1613 played an important role in driving cosmic reionization \citep[e.g.,][]{boylankolchin2015}.

However, there are tensions between the models and data at lower stellar masses. For example, at intermediate masses ($M_{\star} \sim 10^6 - 10^7 M_{\odot}$), Leo~A and DDO~210 are inconsistent with the simulations even when uncertainties are included. The level of disagreement is worse at lower masses, where Phoenix, Tucana, and LGS~3 all exhibit much larger amounts of intermediate and late-time star formation than the simulations predict.  

There are several reasons for these tensions on both the observational and simulations side. Foremost, we are in a regime of small number statistics. Only a handful of isolated dwarf galaxies have `gold standard' SFHs. Similarly, the simulations are only drawn from two realizations of the LG, meaning only $\sim$10 simulated galaxies are in each mass range. Moreover, because the simulations include the entire LG, the mass resolution in the lowest mass dwarfs is coarser compared to zoom-in realizations of individual dwarfs \citep[e.g.,][]{fitts2017}. Numerical tests \citep[e.g.,][]{hopkins2018, garrisonkimmel2019} suggest that dwarf galaxies resolved with fewer than a few 100s of star particles are likely to have their star formation truncated too early. Additionally, as discussed in \citet{garrisonkimmel2019}, the UV background model used \citep[][]{fauchergigure2009} peaks too early compared to current constraints. An updated model \citep[][]{fauchergiguere2019} is currently being implemented but results are not yet available for this analysis. 

Overall, comparisons with the FIRE simulations have generally positive results, with some tensions that warrant further exploration. On one hand, WLM and IC~1613 suggest that simulations for more massive systems are reasonably accurate, modulo our understanding of the very first epochs. This is encouraging, as \citet{skillman2014} noted that many simulations still struggled to reproduce roughly constant SFHs dwarf galaxies. On the other hand, the disagreement at lower masses may indicate that simulations are not accurately reproducing the lowest-mass systems, we do not fully understand our selection effects / systematics, or both. However, before we can drawn any strong conclusions, several competing effects (e.g., the adopted UV background model, small number statistics) must be explored in more detail.  

In terms of selection effects, a galaxy's isolation is a particularly challenging selection effect to quantify. For example, Cetus and Tucana are quenched galaxies located large distances from massive host galaxies, a rarity in the local Universe \citep[e.g.,][]{geha2012}. \citet{teyssier2012} suggest that these two systems likely interacted with the Milky Way in the early Universe, implying that present day isolation does not entirely mitigate the effects of environment. Moreover, galaxies such as Phoenix and LGS~3 are located at intermediate distances from the Milky Way and M31, respectively, and could be considered `backsplash' galaxies, i.e., their evolution may have been affected by weak interactions with their host galaxy \citep[e.g.,][]{ gill2005}. Even if such interactions do not substantially affect a galaxy's stellar populations \citep[e.g., via tidal stripping;][]{knebe2011}, it can strip gas, thereby reducing fuel for future star formation.  

In principle, such effects are included in the simulated SFHs from the FIRE suite we have plotted, as they are selected to be from LG-like environments. However, this also means that the simulations must correctly model galaxy-scale interactions as well as internal processes. It may be that the underlying physics is close but not entirely correct. The use of datasets, such as the detailed SFHs of dwarf galaxies, can be useful to further refine models.

Additionally, it is important to consider HST field location and size in comparing measured and simulated SFHs. As highlighted by \citet{graus2019}, fields that are located within $1 r_h$ are typically biased toward younger ages, relative to the true global SFH, while field located outside $1 r_h$ are biased old.  
For Cetus, Tucana, LGS 3, and Leo A, ACS covers a large fraction of the area around the half-light radius making the bias in mean age fairly modest. However, in galaxies with larger angular sizes such as IC~1613, Phoenix, and WLM, HST covers smaller portions of the galaxies primarily within the half-light radius, potentially biasing the mean age young by $\sim 1$~Gyr \citep{graus2019}. Accordingly, comparisons of SFHs in real and simulated dwarf galaxies must be treated with appropriate caution.  
 
\subsection{Radial Gradients in WLM}
\label{sec:gradients}
 
Comparing SFHs of the ACS and UVIS field reveals a clear age gradient. As summarized in Table \ref{tab:sfh_summary}, the outer and inner fields of WLM formed 50\% of their stellar mass by $\sim 10$~Gyr ago ($z\sim2$) and $\sim 5$~Gyr ago, respectively. This implies a very steep age gradient since the centers of the fields are only $\sim$0.4~kpc apart. Several of the LCID galaxies appear to have gradual and smooth age gradients \citep[e.g.,][]{hidalgo2009, monelli2010b, monelli2010b}, in contrast to what we find for WLM. However, WLM is a more massive system with a disk-like morphology.  

In principle, WLM represents a perfect test bed for the effects of stellar feedback on low-mass galaxy formation. For example, FIRE simulations analyzed by \citet{elbadry2016} find that stellar feedback can take stars initially formed in the central regions and drive them to large radii over cosmic timescales. Interestingly, steeper age gradients are associated with stronger feedback, and in turn, dark matter halos with larger cores in their central regions.  

Using a combination of spectroscopy, photometry, HI observations and dynamic modeling \citet{leaman2012} found a similar story: WLM's evolution appears to be strongly influenced by internal feedback. Interestingly, they found a very weak metallicity gradient of $-0.04$~dex/kpc, which agrees with the metallicites recovered from our CMD fits. Thus, it seems that feedback could induce a steep age gradient, but only a weak metallicity gradient.

As pointed out in \citet{graus2019}, there are biases of up to $\sim 1$~Gyr in the values of $\tau_{50}$ based on the relative radii of the HST fields. However, folding this biases into our analysis is challenging, given that the \citet{graus2019} work uses azimuthally averaged values. Our outer field is along WLM's minor axis, and it remains unclear how SFHs differ along the major and minor axes at a fixed radius. The implications of spatial gradients in WLM is a topic of great interest, and one that we will follow up with in another paper in this series.

\subsection{Toward Globally Representative SFHs of Isolated Dwarf Galaxies}
\label{sec:future}

One challenge in interpreting the SFHs of isolated dwarf galaxies with HST is the effect of a field-of-view (FOV) that is much smaller than the full extent of a galaxy. In almost all cases, HST has targeted fields within or near the half-light radius of isolated dwarfs, largely to provide enough stars for a robust SFH measurement. This placement can introduce biases into the recovered SFHs relative to the true global SFHs \citep[e.g.,][]{graus2019}. For example, SFHs fields located exclusively within the half-light radius may be biased young by up for a few Gyr.

Going forward, mitigating this bias will require coordinated effort among major facilities. The high angular resolution capabilities of HST, and soon JWST, are needed to overcome crowding limitations to reach the oldest MSTO in a galaxy's central region. However, the small fields of these systems are not ideal for the outer regions, which are sparsely populated and subtend large angular areas.  

WFIRST is an excellent solution for measuring SFHs of the outer regions in isolated dwarfs.  Though its angular resolution is twice as coarse as HST in the optical and JWST in the near-IR, its large field-of-view and high throughput are perfect for obtaining deep CMDs of the extended and less crowded halo populations \citep[e.g.,][]{williams2019}. Fully leveraging the complementary nature of these telescopes will require a strategic and coordinated effort across missions.

\section{Conclusions}

We have measured and analyzed the SFH of isolated LG dwarf galaxy WLM based on HST imaging of two fields that reach the oldest MSTO. We find that:

\begin{itemize}
    \item The SFH of the inner ACS field ($0.5 r_h$) is constant or slightly rising toward the present. Prior to 12.5 Gyr ago, this field formed 20\% of its stellar mass, with 50\% formed within the most recent 5~Gyr.  \\

    \item The SFH of the outer UVIS field ($0.7 r_h$) resembles a declining $\tau$-model. This field formed 40\% of its stellar mass by 12.5 Gyr ago, 50\% by 10~Gyr ago, and 90\% by 7~Gyr ago. \\
    
    \item The chemical enrichment of WLM from the CMDs appears consistent with stellar metallicites of RGB stars reported in \citet{leaman2013}. \\
    
    \item We compare the SFHs of WLM and other isolated dwarf galaxies with SFHs of isolated FIRE dwarf galaxy simulations. We find good agreement for galaxies with $M_{\star} > 10^7 M_{\odot}$ and not as good agreement for galaxies with $M_{\star} < 10^7 M_{\odot}$. Small numbers of real galaxies and selection effects make it challenging to draw definitive conclusions. \\
    
    \item The SFHs of the inner and outer field show that WLM has a steep age gradient. The median age changes by 5~Gyr over a distance of only 0.4~kpc. This may imply strong feedback and/or the presence of a prominent dark matter core in WLM. However, because the outer field is along the minor axis, it is unclear if it is representative of the true SFH of WLM at that fixed radius. \\
    
    \item The placement and size of the HST field of view can bias the measured SFHs relative to the true global SFH. Mitigating such biases will require a coordinated effort between HST/JWST to cover the crowded central regions and WFIRST to coverage the sparser, more expansive outer regions.

\end{itemize}

\section*{Acknowledgements}

The authors thank Shea Garrison-Kimmel and Andrew Graus for sharing results from their dwarf galaxy simulations and Piero Madau and Peter Behroozi for helpful comments during the editing process. Support for this work was provided by NASA through grants HST-GO-13768, HST-GO-15006, and JWST-ERS-1334 from the Space Telescope Science Institute, which is operated by AURA, Inc., under NASA contract NAS5-26555. SMA is supported by the National Science Foundation Graduate Research Fellowship under Grant DGE 1752814. DRW acknowledges support from an Alfred P. Sloan Fellowship and an Alexander von Humboldt Fellowship. MBK acknowledges support from NSF grant AST-1517226 and CAREER grant AST-1752913 and from NASA grants NNX17AG29G and HST-AR-14282, HST-AR-14554, and HST-GO-14191 from the Space Telescope Science Institute, which is operated by AURA, Inc., under NASA contract NAS5-26555. This research has made use of NASA's Astrophysics Data System Bibliographic Services and the NASA/IPAC Extragalactic Database (NED), which is operated by the Jet Propulsion Laboratory, California Institute of Technology, under contract with the National Aeronautics and Space Administration.



\bibliographystyle{mnras}
\bibliography{main.bbl}

\begin{thebibliography}{}
\makeatletter
\relax
\def\mn@urlcharsother{\let\do\@makeother \do\$\do\&\do\#\do\^\do\_\do\%\do\~}
\def\mn@doi{\begingroup\mn@urlcharsother \@ifnextchar [ {\mn@doi@}
  {\mn@doi@[]}}
\def\mn@doi@[#1]#2{\def\@tempa{#1}\ifx\@tempa\@empty \href
  {http://dx.doi.org/#2} {doi:#2}\else \href {http://dx.doi.org/#2} {#1}\fi
  \endgroup}
\def\mn@eprint#1#2{\mn@eprint@#1:#2::\@nil}
\def\mn@eprint@arXiv#1{\href {http://arxiv.org/abs/#1} {{\tt arXiv:#1}}}
\def\mn@eprint@dblp#1{\href {http://dblp.uni-trier.de/rec/bibtex/#1.xml}
  {dblp:#1}}
\def\mn@eprint@#1:#2:#3:#4\@nil{\def\@tempa {#1}\def\@tempb {#2}\def\@tempc
  {#3}\ifx \@tempc \@empty \let \@tempc \@tempb \let \@tempb \@tempa \fi \ifx
  \@tempb \@empty \def\@tempb {arXiv}\fi \@ifundefined
  {mn@eprint@\@tempb}{\@tempb:\@tempc}{\expandafter \expandafter \csname
  mn@eprint@\@tempb\endcsname \expandafter{\@tempc}}}

\bibitem[\protect\citeauthoryear{{Aparicio} \& {Hidalgo}}{{Aparicio} \&
  {Hidalgo}}{2009}]{aparicio2009}
{Aparicio} A.,  {Hidalgo} S.~L.,  2009, \mn@doi [\aj]
  {10.1088/0004-6256/138/2/558}, \href
  {http://adsabs.harvard.edu/abs/2009AJ....138..558A} {138, 558}

\bibitem[\protect\citeauthoryear{{Behroozi}, {Wechsler}  \&
  {Conroy}}{{Behroozi} et~al.}{2013}]{behroozi2013}
{Behroozi} P.~S.,  {Wechsler} R.~H.,   {Conroy} C.,  2013, \mn@doi [\apj]
  {10.1088/0004-637X/770/1/57}, \href
  {http://adsabs.harvard.edu/abs/2013ApJ...770...57B} {770, 57}

\bibitem[\protect\citeauthoryear{{Boylan-Kolchin}, {Weisz}, {Johnson},
  {Bullock}, {Conroy}  \& {Fitts}}{{Boylan-Kolchin}
  et~al.}{2015}]{boylankolchin2015}
{Boylan-Kolchin} M.,  {Weisz} D.~R.,  {Johnson} B.~D.,  {Bullock} J.~S.,
  {Conroy} C.,   {Fitts} A.,  2015, \mn@doi [\mnras] {10.1093/mnras/stv1736},
  \href {http://adsabs.harvard.edu/abs/2015MNRAS.453.1503B} {453, 1503}

\bibitem[\protect\citeauthoryear{{Bressan}, {Marigo}, {Girardi}, {Salasnich},
  {Dal Cero}, {Rubele}  \& {Nanni}}{{Bressan} et~al.}{2012}]{bressan2012}
{Bressan} A.,  {Marigo} P.,  {Girardi} L.,  {Salasnich} B.,  {Dal Cero} C.,
  {Rubele} S.,   {Nanni} A.,  2012, \mn@doi [\mnras]
  {10.1111/j.1365-2966.2012.21948.x}, \href
  {http://adsabs.harvard.edu/abs/2012MNRAS.427..127B} {427, 127}

\bibitem[\protect\citeauthoryear{{Buck}, {Macci{\`o}}, {Dutton}, {Obreja}  \&
  {Frings}}{{Buck} et~al.}{2019}]{buck2019}
{Buck} T.,  {Macci{\`o}} A.~V.,  {Dutton} A.~A.,  {Obreja} A.,   {Frings} J.,
  2019, \mn@doi [\mnras] {10.1093/mnras/sty2913}, \href
  {https://ui.adsabs.harvard.edu/abs/2019MNRAS.483.1314B} {483, 1314}

\bibitem[\protect\citeauthoryear{{Choi}, {Dotter}, {Conroy}, {Cantiello},
  {Paxton}  \& {Johnson}}{{Choi} et~al.}{2016}]{choi2016}
{Choi} J.,  {Dotter} A.,  {Conroy} C.,  {Cantiello} M.,  {Paxton} B.,
  {Johnson} B.~D.,  2016, \mn@doi [\apj] {10.3847/0004-637X/823/2/102}, \href
  {http://adsabs.harvard.edu/abs/2016ApJ...823..102C} {823, 102}

\bibitem[\protect\citeauthoryear{{Christensen}, {Dav{\'e}}, {Governato},
  {Pontzen}, {Brooks}, {Munshi}, {Quinn}  \& {Wadsley}}{{Christensen}
  et~al.}{2016}]{christensen2016}
{Christensen} C.~R.,  {Dav{\'e}} R.,  {Governato} F.,  {Pontzen} A.,  {Brooks}
  A.,  {Munshi} F.,  {Quinn} T.,   {Wadsley} J.,  2016, \mn@doi [\apj]
  {10.3847/0004-637X/824/1/57}, \href
  {https://ui.adsabs.harvard.edu/abs/2016ApJ...824...57C} {824, 57}

\bibitem[\protect\citeauthoryear{{Cole} et~al.,}{{Cole}
  et~al.}{2007}]{cole2007}
{Cole} A.~A.,  et~al., 2007, \mn@doi [\apjl] {10.1086/516711}, \href
  {http://adsabs.harvard.edu/abs/2007ApJ...659L..17C} {659, L17}

\bibitem[\protect\citeauthoryear{{Cole}, {Weisz}, {Dolphin}, {Skillman},
  {McConnachie}, {Brooks}  \& {Leaman}}{{Cole} et~al.}{2014}]{cole2014}
{Cole} A.~A.,  {Weisz} D.~R.,  {Dolphin} A.~E.,  {Skillman} E.~D.,
  {McConnachie} A.~W.,  {Brooks} A.~M.,   {Leaman} R.,  2014, \mn@doi [\apj]
  {10.1088/0004-637X/795/1/54}, \href
  {http://adsabs.harvard.edu/abs/2014ApJ...795...54C} {795, 54}

\bibitem[\protect\citeauthoryear{{D'Onghia}, {Besla}, {Cox}  \&
  {Hernquist}}{{D'Onghia} et~al.}{2009}]{donghia2009}
{D'Onghia} E.,  {Besla} G.,  {Cox} T.~J.,   {Hernquist} L.,  2009, \mn@doi
  [\nat] {10.1038/nature08215}, \href
  {http://adsabs.harvard.edu/abs/2009Natur.460..605D} {460, 605}

\bibitem[\protect\citeauthoryear{{Dalcanton} et~al.,}{{Dalcanton}
  et~al.}{2009}]{dalcanton2009}
{Dalcanton} J.~J.,  et~al., 2009, \mn@doi [\apjs] {10.1088/0067-0049/183/1/67},
  \href {http://adsabs.harvard.edu/abs/2009ApJS..183...67D} {183, 67}

\bibitem[\protect\citeauthoryear{{Dolphin}}{{Dolphin}}{2000a}]{dolphin2000b}
{Dolphin} A.~E.,  2000a, \mn@doi [\pasp] {10.1086/316630}, \href
  {http://adsabs.harvard.edu/abs/2000PASP..112.1383D} {112, 1383}

\bibitem[\protect\citeauthoryear{{Dolphin}}{{Dolphin}}{2000b}]{dolphin2000a}
{Dolphin} A.~E.,  2000b, \mn@doi [\apj] {10.1086/308512}, \href
  {http://adsabs.harvard.edu/abs/2000ApJ...531..804D} {531, 804}

\bibitem[\protect\citeauthoryear{{Dolphin}}{{Dolphin}}{2002}]{dolphin2002}
{Dolphin} A.~E.,  2002, \mn@doi [MNRAS] {10.1046/j.1365-8711.2002.05271.x},
  \href {http://adsabs.harvard.edu/abs/2002MNRAS.332...91D} {332, 91}

\bibitem[\protect\citeauthoryear{{Dolphin}}{{Dolphin}}{2012}]{dolphin2012}
{Dolphin} A.~E.,  2012, \mn@doi [\apj] {10.1088/0004-637X/751/1/60}, \href
  {http://adsabs.harvard.edu/abs/2012ApJ...751...60D} {751, 60}

\bibitem[\protect\citeauthoryear{{Dolphin}}{{Dolphin}}{2013}]{dolphin2013}
{Dolphin} A.~E.,  2013, \mn@doi [\apj] {10.1088/0004-637X/775/1/76}, \href
  {http://adsabs.harvard.edu/abs/2013ApJ...775...76D} {775, 76}

\bibitem[\protect\citeauthoryear{{Dolphin} et~al.,}{{Dolphin}
  et~al.}{2003}]{dolphin2003}
{Dolphin} A.~E.,  et~al., 2003, \mn@doi [\aj] {10.1086/375761}, \href
  {http://adsabs.harvard.edu/abs/2003AJ....126..187D} {126, 187}

\bibitem[\protect\citeauthoryear{{Duane}, {Kennedy}, {Pendleton}  \&
  {Roweth}}{{Duane} et~al.}{1987}]{duane1987}
{Duane} S.,  {Kennedy} A.~D.,  {Pendleton} B.~J.,   {Roweth} D.,  1987, \mn@doi
  [Physics Letters B] {10.1016/0370-2693(87)91197-X}, \href
  {http://adsabs.harvard.edu/abs/1987PhLB..195..216D} {195, 216}

\bibitem[\protect\citeauthoryear{{El-Badry}, {Wetzel}, {Geha}, {Hopkins},
  {Kere{\v s}}, {Chan}  \& {Faucher-Gigu{\`e}re}}{{El-Badry}
  et~al.}{2016}]{elbadry2016}
{El-Badry} K.,  {Wetzel} A.,  {Geha} M.,  {Hopkins} P.~F.,  {Kere{\v s}} D.,
  {Chan} T.~K.,   {Faucher-Gigu{\`e}re} C.-A.,  2016, \mn@doi [\apj]
  {10.3847/0004-637X/820/2/131}, \href
  {http://adsabs.harvard.edu/abs/2016ApJ...820..131E} {820, 131}

\bibitem[\protect\citeauthoryear{{El-Badry}, {Weisz}  \& {Quataert}}{{El-Badry}
  et~al.}{2017}]{elbadry2017}
{El-Badry} K.,  {Weisz} D.~R.,   {Quataert} E.,  2017, preprint, \href
  {http://adsabs.harvard.edu/abs/2017arXiv170102347E} {} (\mn@eprint {arXiv}
  {1701.02347})

\bibitem[\protect\citeauthoryear{{Faucher-Giguere} \& {-A.}}{{Faucher-Giguere}
  \& {-A.}}{2019}]{fauchergiguere2019}
{Faucher-Giguere} {-A.} C.,  2019, arXiv e-prints, \href
  {https://ui.adsabs.harvard.edu/abs/2019arXiv190308657F} {p. arXiv:1903.08657}

\bibitem[\protect\citeauthoryear{{Faucher-Gigu{\`e}re}, {Lidz}, {Zaldarriaga}
  \& {Hernquist}}{{Faucher-Gigu{\`e}re} et~al.}{2009}]{fauchergigure2009}
{Faucher-Gigu{\`e}re} C.-A.,  {Lidz} A.,  {Zaldarriaga} M.,   {Hernquist} L.,
  2009, \mn@doi [\apj] {10.1088/0004-637X/703/2/1416}, \href
  {https://ui.adsabs.harvard.edu/abs/2009ApJ...703.1416F} {703, 1416}

\bibitem[\protect\citeauthoryear{{Ferraro}, {Fusi Pecci}, {Tosi}  \&
  {Buonanno}}{{Ferraro} et~al.}{1989}]{ferraro1989}
{Ferraro} F.~R.,  {Fusi Pecci} F.,  {Tosi} M.,   {Buonanno} R.,  1989, \mnras,
  \href {http://adsabs.harvard.edu/abs/1989MNRAS.241..433F} {241, 433}

\bibitem[\protect\citeauthoryear{{Fillingham}, {Cooper}, {Boylan-Kolchin},
  {Bullock}, {Garrison-Kimmel}  \& {Wheeler}}{{Fillingham}
  et~al.}{2018}]{fillingham2018}
{Fillingham} S.~P.,  {Cooper} M.~C.,  {Boylan-Kolchin} M.,  {Bullock} J.~S.,
  {Garrison-Kimmel} S.,   {Wheeler} C.,  2018, \mn@doi [\mnras]
  {10.1093/mnras/sty958}, \href
  {http://adsabs.harvard.edu/abs/2018MNRAS.477.4491F} {477, 4491}

\bibitem[\protect\citeauthoryear{{Fitts} et~al.,}{{Fitts}
  et~al.}{2017}]{fitts2017}
{Fitts} A.,  et~al., 2017, \mn@doi [\mnras] {10.1093/mnras/stx1757}, \href
  {http://adsabs.harvard.edu/abs/2017MNRAS.471.3547F} {471, 3547}

\bibitem[\protect\citeauthoryear{{Fitts} et~al.,}{{Fitts}
  et~al.}{2018}]{fitts2018}
{Fitts} A.,  et~al., 2018, \mn@doi [\mnras] {10.1093/mnras/sty1488}, \href
  {http://adsabs.harvard.edu/abs/2018MNRAS.479..319F} {479, 319}

\bibitem[\protect\citeauthoryear{{Gallart}, {Zoccali}  \& {Aparicio}}{{Gallart}
  et~al.}{2005}]{Gallart:2005qy}
{Gallart} C.,  {Zoccali} M.,   {Aparicio} A.,  2005, \mn@doi [\araa]
  {10.1146/annurev.astro.43.072103.150608}, \href
  {http://adsabs.harvard.edu/abs/2005ARA%26A..43..387G} {43, 387}

\bibitem[\protect\citeauthoryear{{Gallart} et~al.,}{{Gallart}
  et~al.}{2015}]{gallart2015}
{Gallart} C.,  et~al., 2015, \mn@doi [\apjl] {10.1088/2041-8205/811/2/L18},
  \href {http://adsabs.harvard.edu/abs/2015ApJ...811L..18G} {811, L18}

\bibitem[\protect\citeauthoryear{{Gao}, {Springel}  \& {White}}{{Gao}
  et~al.}{2005}]{gao2005}
{Gao} L.,  {Springel} V.,   {White} S. D.~M.,  2005, \mn@doi [\mnras]
  {10.1111/j.1745-3933.2005.00084.x}, \href
  {https://ui.adsabs.harvard.edu/abs/2005MNRAS.363L..66G} {363, L66}

\bibitem[\protect\citeauthoryear{{Garrison-Kimmel} et~al.,}{{Garrison-Kimmel}
  et~al.}{2019}]{garrisonkimmel2019}
{Garrison-Kimmel} S.,  et~al., 2019, arXiv e-prints, \href
  {https://ui.adsabs.harvard.edu/\#abs/2019arXiv190310515G} {p.
  arXiv:1903.10515}

\bibitem[\protect\citeauthoryear{{Geha}, {Blanton}, {Yan}  \& {Tinker}}{{Geha}
  et~al.}{2012}]{geha2012}
{Geha} M.,  {Blanton} M.~R.,  {Yan} R.,   {Tinker} J.~L.,  2012, \mn@doi [\apj]
  {10.1088/0004-637X/757/1/85}, \href
  {http://adsabs.harvard.edu/abs/2012ApJ...757...85G} {757, 85}

\bibitem[\protect\citeauthoryear{{Gieren} et~al.,}{{Gieren}
  et~al.}{2008}]{gieren2008}
{Gieren} W.,  et~al., 2008, \mn@doi [\apj] {10.1086/589994}, \href
  {http://adsabs.harvard.edu/abs/2008ApJ...683..611G} {683, 611}

\bibitem[\protect\citeauthoryear{{Gill}, {Knebe}  \& {Gibson}}{{Gill}
  et~al.}{2005}]{gill2005}
{Gill} S. P.~D.,  {Knebe} A.,   {Gibson} B.~K.,  2005, \mn@doi [\mnras]
  {10.1111/j.1365-2966.2004.08562.x}, \href
  {https://ui.adsabs.harvard.edu/abs/2005MNRAS.356.1327G} {356, 1327}

\bibitem[\protect\citeauthoryear{{Girardi} et~al.,}{{Girardi}
  et~al.}{2010}]{girardi2010}
{Girardi} L.,  et~al., 2010, \mn@doi [\apj] {10.1088/0004-637X/724/2/1030},
  \href {http://adsabs.harvard.edu/abs/2010ApJ...724.1030G} {724, 1030}

\bibitem[\protect\citeauthoryear{{Gonz{\'a}lez-Samaniego}, {Bullock},
  {Boylan-Kolchin}, {Fitts}, {Elbert}, {Hopkins}, {Kere{\v s}}  \&
  {Faucher-Gigu{\`e}re}}{{Gonz{\'a}lez-Samaniego}
  et~al.}{2017}]{gonzalezsamaniego2017}
{Gonz{\'a}lez-Samaniego} A.,  {Bullock} J.~S.,  {Boylan-Kolchin} M.,  {Fitts}
  A.,  {Elbert} O.~D.,  {Hopkins} P.~F.,  {Kere{\v s}} D.,
  {Faucher-Gigu{\`e}re} C.-A.,  2017, \mn@doi [\mnras] {10.1093/mnras/stx2322},
  \href {http://adsabs.harvard.edu/abs/2017MNRAS.472.4786G} {472, 4786}

\bibitem[\protect\citeauthoryear{{Governato} et~al.,}{{Governato}
  et~al.}{2010}]{governato2010}
{Governato} F.,  et~al., 2010, \mn@doi [\nat] {10.1038/nature08640}, \href
  {http://adsabs.harvard.edu/abs/2010Natur.463..203G} {463, 203}

\bibitem[\protect\citeauthoryear{{Graus} et~al.,}{{Graus}
  et~al.}{2019}]{graus2019}
{Graus} A.~S.,  et~al., 2019, arXiv e-prints, \href
  {http://adsabs.harvard.edu/abs/2019arXiv190105487G} {}

\bibitem[\protect\citeauthoryear{{Hidalgo}, {Aparicio},
  {Mart{\'{\i}}nez-Delgado}  \& {Gallart}}{{Hidalgo}
  et~al.}{2009}]{hidalgo2009}
{Hidalgo} S.~L.,  {Aparicio} A.,  {Mart{\'{\i}}nez-Delgado} D.,   {Gallart} C.,
   2009, \mn@doi [\apj] {10.1088/0004-637X/705/1/704}, \href
  {http://adsabs.harvard.edu/abs/2009ApJ...705..704H} {705, 704}

\bibitem[\protect\citeauthoryear{{Hidalgo} et~al.,}{{Hidalgo}
  et~al.}{2018}]{hidalgo2018}
{Hidalgo} S.~L.,  et~al., 2018, \mn@doi [\apj] {10.3847/1538-4357/aab158},
  \href {https://ui.adsabs.harvard.edu/abs/2018ApJ...856..125H} {856, 125}

\bibitem[\protect\citeauthoryear{{Hodge} \& {Miller}}{{Hodge} \&
  {Miller}}{1995}]{hodge1995}
{Hodge} P.,  {Miller} B.~W.,  1995, \mn@doi [\apj] {10.1086/176209}, \href
  {http://adsabs.harvard.edu/abs/1995ApJ...451..176H} {451, 176}

\bibitem[\protect\citeauthoryear{{Holtzman}, {Afonso}  \& {Dolphin}}{{Holtzman}
  et~al.}{2006}]{holtzman2006}
{Holtzman} J.~A.,  {Afonso} C.,   {Dolphin} A.,  2006, \mn@doi [\apjs]
  {10.1086/507074}, \href {http://adsabs.harvard.edu/abs/2006ApJS..166..534H}
  {166, 534}

\bibitem[\protect\citeauthoryear{{Hopkins}, {Kere{\v s}}, {O{\~n}orbe},
  {Faucher-Gigu{\`e}re}, {Quataert}, {Murray}  \& {Bullock}}{{Hopkins}
  et~al.}{2014}]{hopkins2014}
{Hopkins} P.~F.,  {Kere{\v s}} D.,  {O{\~n}orbe} J.,  {Faucher-Gigu{\`e}re}
  C.-A.,  {Quataert} E.,  {Murray} N.,   {Bullock} J.~S.,  2014, \mn@doi
  [\mnras] {10.1093/mnras/stu1738}, \href
  {http://adsabs.harvard.edu/abs/2014MNRAS.445..581H} {445, 581}

\bibitem[\protect\citeauthoryear{{Hopkins} et~al.,}{{Hopkins}
  et~al.}{2018}]{hopkins2018}
{Hopkins} P.~F.,  et~al., 2018, \mn@doi [\mnras] {10.1093/mnras/sty1690}, \href
  {https://ui.adsabs.harvard.edu/abs/2018MNRAS.480..800H} {480, 800}

\bibitem[\protect\citeauthoryear{{Karachentsev}, {Makarov}  \&
  {Kaisina}}{{Karachentsev} et~al.}{2013}]{karachentsev2013}
{Karachentsev} I.~D.,  {Makarov} D.~I.,   {Kaisina} E.~I.,  2013, \mn@doi [\aj]
  {10.1088/0004-6256/145/4/101}, \href
  {http://adsabs.harvard.edu/abs/2013AJ....145..101K} {145, 101}

\bibitem[\protect\citeauthoryear{{Kirby}, {Cohen}, {Simon}, {Guhathakurta},
  {Thygesen}  \& {Duggan}}{{Kirby} et~al.}{2017}]{kirby2017}
{Kirby} E.~N.,  {Cohen} J.~G.,  {Simon} J.~D.,  {Guhathakurta} P.,  {Thygesen}
  A.~O.,   {Duggan} G.~E.,  2017, \mn@doi [\apj] {10.3847/1538-4357/aa6570},
  \href {https://ui.adsabs.harvard.edu/\#abs/2017ApJ...838...83K} {838, 83}

\bibitem[\protect\citeauthoryear{{Knebe}, {Libeskind}, {Knollmann},
  {Martinez-Vaquero}, {Yepes}, {Gottl{\"o}ber}  \& {Hoffman}}{{Knebe}
  et~al.}{2011}]{knebe2011}
{Knebe} A.,  {Libeskind} N.~I.,  {Knollmann} S.~R.,  {Martinez-Vaquero} L.~A.,
  {Yepes} G.,  {Gottl{\"o}ber} S.,   {Hoffman} Y.,  2011, \mn@doi [\mnras]
  {10.1111/j.1365-2966.2010.17924.x}, \href
  {https://ui.adsabs.harvard.edu/abs/2011MNRAS.412..529K} {412, 529}

\bibitem[\protect\citeauthoryear{{Kroupa}}{{Kroupa}}{2001}]{kroupa2001}
{Kroupa} P.,  2001, \mn@doi [\mnras] {10.1046/j.1365-8711.2001.04022.x}, \href
  {http://adsabs.harvard.edu/abs/2001MNRAS.322..231K} {322, 231}

\bibitem[\protect\citeauthoryear{{Leaman} et~al.,}{{Leaman}
  et~al.}{2012}]{leaman2012}
{Leaman} R.,  et~al., 2012, \mn@doi [\apj] {10.1088/0004-637X/750/1/33}, \href
  {http://adsabs.harvard.edu/abs/2012ApJ...750...33L} {750, 33}

\bibitem[\protect\citeauthoryear{{Leaman} et~al.,}{{Leaman}
  et~al.}{2013}]{leaman2013}
{Leaman} R.,  et~al., 2013, \mn@doi [\apj] {10.1088/0004-637X/767/2/131}, \href
  {http://adsabs.harvard.edu/abs/2013ApJ...767..131L} {767, 131}

\bibitem[\protect\citeauthoryear{{Lee}, {Skillman}  \& {Venn}}{{Lee}
  et~al.}{2005}]{lee2005}
{Lee} H.,  {Skillman} E.~D.,   {Venn} K.~A.,  2005, \mn@doi [\apj]
  {10.1086/427019}, \href {http://adsabs.harvard.edu/abs/2005ApJ...620..223L}
  {620, 223}

\bibitem[\protect\citeauthoryear{{Makarov}, {Makarova}, {Rizzi}, {Tully},
  {Dolphin}, {Sakai}  \& {Shaya}}{{Makarov} et~al.}{2006}]{makarov2006}
{Makarov} D.,  {Makarova} L.,  {Rizzi} L.,  {Tully} R.~B.,  {Dolphin} A.~E.,
  {Sakai} S.,   {Shaya} E.~J.,  2006, \mn@doi [\aj] {10.1086/508925}, \href
  {http://adsabs.harvard.edu/abs/2006AJ....132.2729M} {132, 2729}

\bibitem[\protect\citeauthoryear{{McConnachie}}{{McConnachie}}{2012}]{mcconnachie2012}
{McConnachie} A.~W.,  2012, \mn@doi [\aj] {10.1088/0004-6256/144/1/4}, \href
  {http://adsabs.harvard.edu/abs/2012AJ....144....4M} {144, 4}

\bibitem[\protect\citeauthoryear{{McQuinn}, {Skillman}, {Dalcanton}, {Dolphin},
  {Holtzman}, {Weisz}  \& {Williams}}{{McQuinn} et~al.}{2011}]{mcquinn2011}
{McQuinn} K.~B.~W.,  {Skillman} E.~D.,  {Dalcanton} J.~J.,  {Dolphin} A.~E.,
  {Holtzman} J.,  {Weisz} D.~R.,   {Williams} B.~F.,  2011, \mn@doi [\apj]
  {10.1088/0004-637X/740/1/48}, \href
  {http://adsabs.harvard.edu/abs/2011ApJ...740...48M} {740, 48}

\bibitem[\protect\citeauthoryear{{Minniti} \& {Zijlstra}}{{Minniti} \&
  {Zijlstra}}{1996}]{minniti1996}
{Minniti} D.,  {Zijlstra} A.~A.,  1996, \mn@doi [\apjl] {10.1086/310189}, \href
  {http://adsabs.harvard.edu/abs/1996ApJ...467L..13M} {467, L13}

\bibitem[\protect\citeauthoryear{{Minniti} \& {Zijlstra}}{{Minniti} \&
  {Zijlstra}}{1997}]{minniti1997}
{Minniti} D.,  {Zijlstra} A.~A.,  1997, \mn@doi [\aj] {10.1086/118461}, \href
  {http://adsabs.harvard.edu/abs/1997AJ....114..147M} {114, 147}

\bibitem[\protect\citeauthoryear{{Monelli} et~al.,}{{Monelli}
  et~al.}{2010a}]{monelli2010b}
{Monelli} M.,  et~al., 2010a, \mn@doi [\apj] {10.1088/0004-637X/720/2/1225},
  \href {http://adsabs.harvard.edu/abs/2010ApJ...720.1225M} {720, 1225}

\bibitem[\protect\citeauthoryear{{Monelli} et~al.,}{{Monelli}
  et~al.}{2010b}]{monelli2010c}
{Monelli} M.,  et~al., 2010b, \mn@doi [\apj] {10.1088/0004-637X/722/2/1864},
  \href {http://adsabs.harvard.edu/abs/2010ApJ...722.1864M} {722, 1864}

\bibitem[\protect\citeauthoryear{{Rizzi}, {Tully}, {Makarov}, {Makarova},
  {Dolphin}, {Sakai}  \& {Shaya}}{{Rizzi} et~al.}{2007}]{rizzi2007}
{Rizzi} L.,  {Tully} R.~B.,  {Makarov} D.,  {Makarova} L.,  {Dolphin} A.~E.,
  {Sakai} S.,   {Shaya} E.~J.,  2007, \mn@doi [\apj] {10.1086/516566}, \href
  {http://adsabs.harvard.edu/abs/2007ApJ...661..815R} {661, 815}

\bibitem[\protect\citeauthoryear{{Robles} et~al.,}{{Robles}
  et~al.}{2017}]{robles2017}
{Robles} V.~H.,  et~al., 2017, \mn@doi [\mnras] {10.1093/mnras/stx2253}, \href
  {http://adsabs.harvard.edu/abs/2017MNRAS.472.2945R} {472, 2945}

\bibitem[\protect\citeauthoryear{{Rosenfield}, {Marigo}, {Girardi},
  {Dalcanton}, {Bressan}, {Williams}  \& {Dolphin}}{{Rosenfield}
  et~al.}{2016}]{rosenfield2016}
{Rosenfield} P.,  {Marigo} P.,  {Girardi} L.,  {Dalcanton} J.~J.,  {Bressan}
  A.,  {Williams} B.~F.,   {Dolphin} A.,  2016, preprint, \href
  {http://adsabs.harvard.edu/abs/2016arXiv160305283R} {} (\mn@eprint {arXiv}
  {1603.05283})

\bibitem[\protect\citeauthoryear{{Rosenfield} et~al.,}{{Rosenfield}
  et~al.}{2017}]{rosenfield2017}
{Rosenfield} P.,  et~al., 2017, \mn@doi [\apj] {10.3847/1538-4357/aa70a2},
  \href {https://ui.adsabs.harvard.edu/abs/2017ApJ...841...69R} {841, 69}

\bibitem[\protect\citeauthoryear{{Sawala}, {Scannapieco}, {Maio}  \&
  {White}}{{Sawala} et~al.}{2010}]{sawala2010}
{Sawala} T.,  {Scannapieco} C.,  {Maio} U.,   {White} S.,  2010, \mn@doi
  [\mnras] {10.1111/j.1365-2966.2009.16035.x}, \href
  {http://adsabs.harvard.edu/abs/2010MNRAS.402.1599S} {402, 1599}

\bibitem[\protect\citeauthoryear{{Sawala} et~al.,}{{Sawala}
  et~al.}{2016}]{sawala2016}
{Sawala} T.,  et~al., 2016, \mn@doi [\mnras] {10.1093/mnras/stw145}, \href
  {http://adsabs.harvard.edu/abs/2016MNRAS.457.1931S} {457, 1931}

\bibitem[\protect\citeauthoryear{{Schlafly} \& {Finkbeiner}}{{Schlafly} \&
  {Finkbeiner}}{2011}]{schlafly2011}
{Schlafly} E.~F.,  {Finkbeiner} D.~P.,  2011, \mn@doi [\apj]
  {10.1088/0004-637X/737/2/103}, \href
  {http://adsabs.harvard.edu/abs/2011ApJ...737..103S} {737, 103}

\bibitem[\protect\citeauthoryear{{Shen}, {Madau}, {Conroy}, {Governato}  \&
  {Mayer}}{{Shen} et~al.}{2014}]{shen2014}
{Shen} S.,  {Madau} P.,  {Conroy} C.,  {Governato} F.,   {Mayer} L.,  2014,
  \mn@doi [\apj] {10.1088/0004-637X/792/2/99}, \href
  {https://ui.adsabs.harvard.edu/abs/2014ApJ...792...99S} {792, 99}

\bibitem[\protect\citeauthoryear{{Skillman}, {Terlevich}  \&
  {Melnick}}{{Skillman} et~al.}{1989}]{skillman1989}
{Skillman} E.~D.,  {Terlevich} R.,   {Melnick} J.,  1989, \mn@doi [\mnras]
  {10.1093/mnras/240.3.563}, \href
  {http://adsabs.harvard.edu/abs/1989MNRAS.240..563S} {240, 563}

\bibitem[\protect\citeauthoryear{{Skillman} et~al.,}{{Skillman}
  et~al.}{2014}]{skillman2014}
{Skillman} E.~D.,  et~al., 2014, \mn@doi [\apj] {10.1088/0004-637X/786/1/44},
  \href {http://adsabs.harvard.edu/abs/2014ApJ...786...44S} {786, 44}

\bibitem[\protect\citeauthoryear{{Skillman} et~al.,}{{Skillman}
  et~al.}{2017}]{skillman2017}
{Skillman} E.~D.,  et~al., 2017, \mn@doi [\apj] {10.3847/1538-4357/aa60c5},
  \href {http://adsabs.harvard.edu/abs/2017ApJ...837..102S} {837, 102}

\bibitem[\protect\citeauthoryear{{Su} et~al.,}{{Su} et~al.}{2018}]{su2018}
{Su} K.-Y.,  et~al., 2018, \mn@doi [\mnras] {10.1093/mnras/sty1928}, \href
  {http://adsabs.harvard.edu/abs/2018MNRAS.480.1666S} {480, 1666}

\bibitem[\protect\citeauthoryear{{Teyssier}, {Johnston}  \&
  {Kuhlen}}{{Teyssier} et~al.}{2012}]{teyssier2012}
{Teyssier} M.,  {Johnston} K.~V.,   {Kuhlen} M.,  2012, \mn@doi [\mnras]
  {10.1111/j.1365-2966.2012.21793.x}, \href
  {http://adsabs.harvard.edu/abs/2012MNRAS.426.1808T} {426, 1808}

\bibitem[\protect\citeauthoryear{{Tolstoy} et~al.,}{{Tolstoy}
  et~al.}{1998}]{tolstoy1998}
{Tolstoy} E.,  et~al., 1998, \mn@doi [\aj] {10.1086/300515}, \href
  {http://adsabs.harvard.edu/abs/1998AJ....116.1244T} {116, 1244}

\bibitem[\protect\citeauthoryear{{Urbaneja}, {Kudritzki}, {Bresolin},
  {Przybilla}, {Gieren}  \& {Pietrzy{\'n}ski}}{{Urbaneja}
  et~al.}{2008}]{urbaneja2008}
{Urbaneja} M.~A.,  {Kudritzki} R.-P.,  {Bresolin} F.,  {Przybilla} N.,
  {Gieren} W.,   {Pietrzy{\'n}ski} G.,  2008, \mn@doi [\apj] {10.1086/590334},
  \href {http://adsabs.harvard.edu/abs/2008ApJ...684..118U} {684, 118}

\bibitem[\protect\citeauthoryear{{Vogelsberger}, {Zavala}, {Simpson}  \&
  {Jenkins}}{{Vogelsberger} et~al.}{2014}]{vogelsberger2014}
{Vogelsberger} M.,  {Zavala} J.,  {Simpson} C.,   {Jenkins} A.,  2014, \mn@doi
  [\mnras] {10.1093/mnras/stu1713}, \href
  {https://ui.adsabs.harvard.edu/abs/2014MNRAS.444.3684V} {444, 3684}

\bibitem[\protect\citeauthoryear{{Weisz} et~al.,}{{Weisz}
  et~al.}{2011}]{weisz2011a}
{Weisz} D.~R.,  et~al., 2011, \mn@doi [\apj] {10.1088/0004-637X/739/1/5}, \href
  {http://adsabs.harvard.edu/abs/2011ApJ...739....5W} {739, 5}

\bibitem[\protect\citeauthoryear{{Weisz}, {Dolphin}, {Skillman}, {Holtzman},
  {Gilbert}, {Dalcanton}  \& {Williams}}{{Weisz} et~al.}{2014}]{weisz2014a}
{Weisz} D.~R.,  {Dolphin} A.~E.,  {Skillman} E.~D.,  {Holtzman} J.,  {Gilbert}
  K.~M.,  {Dalcanton} J.~J.,   {Williams} B.~F.,  2014, \mn@doi [\apj]
  {10.1088/0004-637X/789/2/147}, \href
  {http://adsabs.harvard.edu/abs/2014ApJ...789..147W} {789, 147}

\bibitem[\protect\citeauthoryear{{Wheeler} et~al.,}{{Wheeler}
  et~al.}{2018}]{wheeler2018}
{Wheeler} C.,  et~al., 2018, arXiv e-prints, \href
  {https://ui.adsabs.harvard.edu/abs/2018arXiv181202749W} {p. arXiv:1812.02749}

\bibitem[\protect\citeauthoryear{{Williams} et~al.,}{{Williams}
  et~al.}{2014}]{williams2014}
{Williams} B.~F.,  et~al., 2014, \mn@doi [\apjs] {10.1088/0067-0049/215/1/9},
  \href {http://adsabs.harvard.edu/abs/2014ApJS..215....9W} {215, 9}

\bibitem[\protect\citeauthoryear{{Williams} et~al.,}{{Williams}
  et~al.}{2019}]{williams2019}
{Williams} B.,  et~al., 2019, in \baas. p.~301

\bibitem[\protect\citeauthoryear{{Wright}, {Brooks}, {Weisz}  \&
  {Christensen}}{{Wright} et~al.}{2019}]{wright2019}
{Wright} A.~C.,  {Brooks} A.~M.,  {Weisz} D.~R.,   {Christensen} C.~R.,  2019,
  \mn@doi [\mnras] {10.1093/mnras/sty2759}, \href
  {https://ui.adsabs.harvard.edu/abs/2019MNRAS.482.1176W} {482, 1176}

\makeatother
\end{thebibliography}



\appendix

\section{Tabulated Star Formation Histories}
\label{sec:appenda}

Tables \ref{tab:sfh_acs} and \ref{tab:sfh_uvis} list the cumulative and absolute SFHs, along with metallicity enrichment, for the ACS and UVIS fields, respectively. The first error listed is the random error, the second error is the total error (random plus systematic). 10 lines are listed in these tables. The full tables are available online.

\begin{table}
\caption{Column (1) log youngest look back time in bin; Column (2) log oldest look back time in bin; Column (3) cumulative SFH; Column (4) Absolute SFH; Column (5) Mean Metallicity.  Note that the value of [M/H] is set to 0 when there is no star formation.}
\label{tab:sfh_acs}
\begin{center}
 
\begin{tabular}{ c|c|c|c|c }
\toprule[1pt]\midrule[0.3pt]
log($t_1$) & log($t_2$) & cSFH & SFR  & [M/H] \\
(yr ago) & (yr ago) &  & (10$^{-3}$ M$_{{\odot}}$ /yr) & (dex) \\
\hline
(1) & (2) & (3) & (4) & (5) \\
\hline
6.6 & 6.7 & 1.00$_{-0.00, 0.00}^{+0.00, 0.00}$ & 0.00$_{-0.00, 0.00}^{+3.66, 3.66}$ & 0.00$_{-0.70, 0.70}^{+0.00, 0.00}$ \\ 
6.7 & 6.8 & 1.00$_{-0.00, 0.00}^{+0.00, 0.00}$ & 0.00$_{-0.00, 0.00}^{+3.71, 3.71}$ & 0.00$_{-0.70, 0.70}^{+0.00, 0.00}$ \\ 
6.8 & 6.9 & 1.00$_{-0.00, 0.00}^{+0.00, 0.00}$ & 0.00$_{-0.00, 0.00}^{+3.91, 3.91}$ & 0.00$_{-0.70, 0.70}^{+0.00, 0.00}$ \\ 
6.9 & 7.0 & 1.00$_{-0.00, 0.00}^{+0.00, 0.00}$ & 4.03$_{-3.47, 4.03}^{+2.33, 2.33}$ & -0.70$_{-0.00, 0.00}^{+0.00, 0.39}$ \\ 
7.0 & 7.1 & 1.00$_{-0.00, 0.00}^{+0.00, 0.00}$ & 6.89$_{-5.54, 5.61}^{+0.05, 1.82}$ & -0.70$_{-0.00, 0.40}^{+0.00, 0.10}$ \\ 
7.1 & 7.2 & 1.00$_{-0.00, 0.00}^{+0.00, 0.00}$ & 0.00$_{-0.00, 0.00}^{+2.07, 2.07}$ & 0.00$_{-0.70, 1.30}^{+0.00, 0.00}$ \\ 
7.2 & 7.3 & 1.00$_{-0.00, 0.00}^{+0.00, 0.00}$ & 0.00$_{-0.00, 0.00}^{+1.16, 1.16}$ & 0.00$_{-0.70, 0.70}^{+0.00, 0.00}$ \\ 
7.3 & 7.4 & 1.00$_{-0.00, 0.00}^{+0.00, 0.00}$ & 0.00$_{-0.00, 0.00}^{+0.82, 0.82}$ & 0.00$_{-0.70, 0.70}^{+0.00, 0.00}$ \\ 
7.4 & 7.5 & 1.00$_{-0.00, 0.00}^{+0.00, 0.00}$ & 0.00$_{-0.00, 0.00}^{+0.65, 0.65}$ & 0.00$_{-0.70, 0.70}^{+0.00, 0.00}$ \\ 
7.5 & 7.6 & 1.00$_{-0.00, 0.00}^{+0.00, 0.00}$ & 5.69$_{-0.99, 1.00}^{+0.28, 3.53}$ & -0.70$_{-0.00, 0.40}^{+0.00, 0.20}$ \\ 
\hline
\end{tabular}
\end{center}
\end{table}

\begin{table}
\caption{Same as Table \ref{tab:sfh_acs} only for the UVIS field.}
\label{tab:sfh_uvis}
\begin{center}
\begin{tabular}{ c|c|c|c|c }
\toprule[1pt]\midrule[0.3pt]
log($t_1$) & log($t_2$) & cSFH & SFR  & [M/H] \\
(yr ago) & (yr ago) &  & (10$^{-3}$ M$_{{\odot}}$ /yr) & (dex) \\
\hline
(1) & (2) & (3) & (4) & (5) \\
\hline
6.6 & 6.7 & 1.00$_{-0.00, 0.00}^{+0.00, 0.00}$ & 0.00$_{-0.00, 0.00}^{+0.07, 0.07}$ & 0.00$_{-0.80, 1.00}^{+0.00, 0.00}$ \\ 
6.7 & 6.8 & 1.00$_{-0.00, 0.00}^{+0.00, 0.00}$ & 0.00$_{-0.00, 0.00}^{+0.05, 0.05}$ & 0.00$_{-0.80, 0.80}^{+0.00, 0.00}$ \\ 
6.8 & 6.9 & 1.00$_{-0.00, 0.00}^{+0.00, 0.00}$ & 0.00$_{-0.00, 0.00}^{+0.04, 0.04}$ & 0.00$_{-0.80, 0.80}^{+0.00, 0.00}$ \\ 
6.9 & 7.0 & 1.00$_{-0.00, 0.00}^{+0.00, 0.00}$ & 0.00$_{-0.00, 0.00}^{+0.03, 0.03}$ & 0.00$_{-0.80, 0.80}^{+0.00, 0.00}$ \\ 
7.0 & 7.1 & 1.00$_{-0.00, 0.00}^{+0.00, 0.00}$ & 0.00$_{-0.00, 0.00}^{+0.03, 0.03}$ & 0.00$_{-0.80, 0.80}^{+0.00, 0.00}$ \\ 
7.1 & 7.2 & 1.00$_{-0.00, 0.00}^{+0.00, 0.00}$ & 0.00$_{-0.00, 0.00}^{+0.02, 0.02}$ & 0.00$_{-0.80, 0.80}^{+0.00, 0.00}$ \\ 
7.2 & 7.3 & 1.00$_{-0.00, 0.00}^{+0.00, 0.00}$ & 0.00$_{-0.00, 0.00}^{+0.02, 0.02}$ & 0.00$_{-0.80, 0.80}^{+0.00, 0.00}$ \\ 
7.3 & 7.4 & 1.00$_{-0.00, 0.00}^{+0.00, 0.00}$ & 0.00$_{-0.00, 0.00}^{+0.01, 0.01}$ & 0.00$_{-0.80, 0.80}^{+0.00, 0.00}$ \\ 
7.4 & 7.5 & 1.00$_{-0.00, 0.00}^{+0.00, 0.00}$ & 0.00$_{-0.00, 0.00}^{+0.01, 0.01}$ & 0.00$_{-0.80, 0.80}^{+0.00, 0.00}$ \\ 
7.5 & 7.6 & 1.00$_{-0.00, 0.00}^{+0.00, 0.00}$ & 0.00$_{-0.00, 0.00}^{+0.01, 0.01}$ & 0.00$_{-0.80, 0.80}^{+0.00, 0.00}$ \\ 

\hline
\end{tabular}
\end{center}
\end{table}

\section{Effects of Stellar Models on the Star Formation History of WLM}
\label{sec:appendb}

Choice of underlying stellar model has been a long standing challenge in translating CMDs to SFHs. CMDs are so information rich that variations in the underlying stellar interior, atmosphere physics (e.g., choice in mixing length value, boundary conditions, nuclear reaction rates) and associated implementation can lead to differences in ages and SFHs.  \citet{Gallart:2005qy} present an in-depth discussion on how adopted physics in given stellar models affect the CMDs they predict. \citet{choi2016} includes similar discussion with newer stellar models.

The effect of stellar models on SFHs measured from CMDs has been explored extensively in the past \citep[e.g.,][]{aparicio2009, weisz2011a, dolphin2012, weisz2014a, skillman2017}. In essence, when a CMD reaches the oldest MSTO with sufficient SNR ($\gtrsim 5-10$), the resulting SFH weakly depends on the choice in stellar models. The impact of stellar models on the SFH becomes larger as the CMD gets shallower. This is because SFHs from shallower CMDs are measured only from evolved stars (e.g., RGB, HB, AGB) whose age sensitivity is poorer than MSTO and sub-giants. Additionally, their exact location on the CMD is more sensitive to choice in underlying stellar model.

\begin{figure}
\centering
	\includegraphics[scale=0.3]{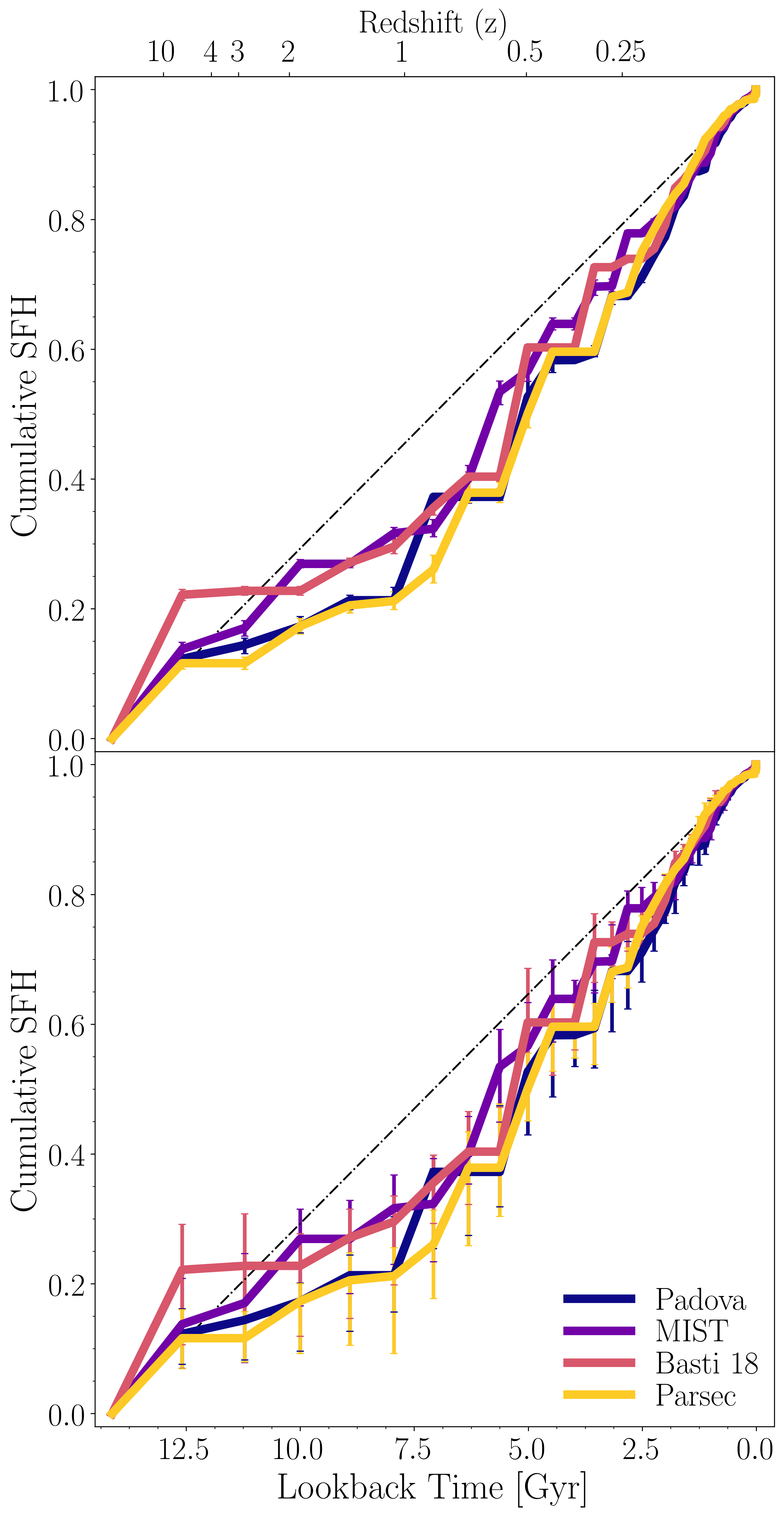}
    \caption{The above plot shows the star formation history for the inner ACS field of WLM using different stellar models. The top panel displays only the random errors, while the bottom panel displays the total error (random plus systematic). Overall, there is good agreement between the different models.}
    \label{fig:acs_sfh_models}
\end{figure}

\begin{figure}
\centering
	\includegraphics[scale=0.3]{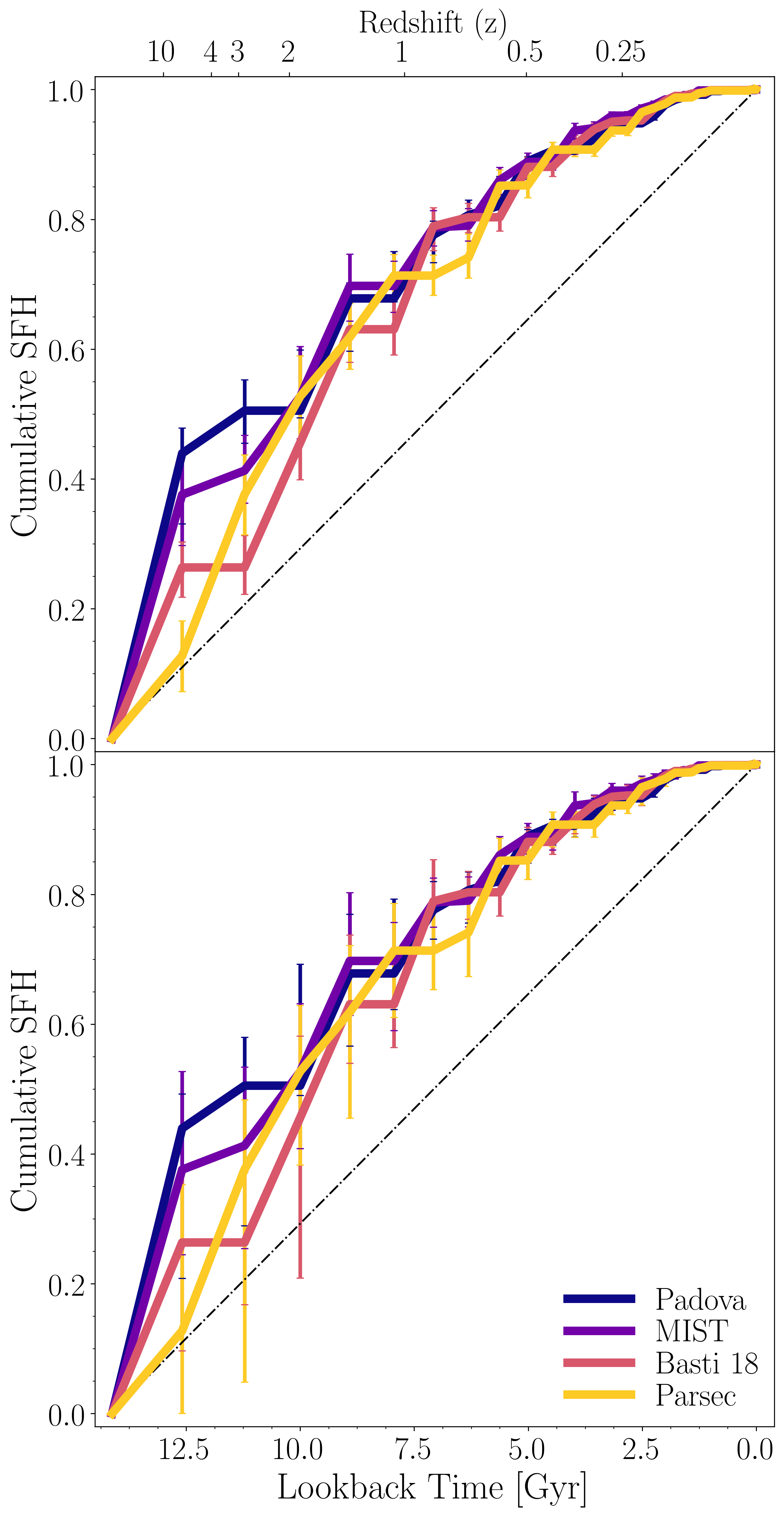}
    \caption{The above plot shows the star formation history for the outer UVIS field of WLM using different stellar models. The top panel displays only the random errors, while the bottom panel displays the total error (random plus systematic). The overall agreement between the different models supports the robustness of our measured star formation history. }
    \label{fig:uvis_sfh_models}
\end{figure}

In the process of measuring the SFH of WLM from our deep HST data, we varied the choice in stellar model, holding all other parameters fixed (e.g., same IMF, age/metallicity binning, distance). The results for the ACS and UVIS fields are shown in Figures \ref{fig:acs_sfh_models} and \ref{fig:uvis_sfh_models}, respectively. The SFHs are color-coded by stellar model: Padova \citep[navy;][]{girardi2010}; MIST \citep[purple;][]{choi2016}; BaSTI  \citep[magenta;][]{hidalgo2018}; PARSEC \citep[yellow;][]{bressan2012}.  

The upper panel in each plot shows the best fit cumulative SFHs for each model with random uncertainties, i.e., computed following \citet{dolphin2013}. In both the ACS and UVIS fields, the SFHs are qualitatively quite similar, which is expected from such deep data. Because the random errors are quite small, they are often in tension at the several sigma level on a strictly statistical basis. This is perhaps useful for diagnosing challenges in the underlying physics \citep[e.g.,][]{rosenfield2016, rosenfield2017}, but it poses a challenge for interpreting the SFH of a galaxy, i.e., which one is correct?  

One solution proposed by \citet{dolphin2012} is to include an error term on the SFH that is meant to encompass plausible variations on the stellar models. More specifically, the procedure is to use a Monte Carlo process to sample variations in stellar models in the $M_{\rm bol}$ - $T_{\rm eff}$ plane, and re-fit the CMD with a slightly perturbed set of models. Repeating this process for many iterations is designed to produce an error estimate that captures the effects of varying stellar models. Implementation and calibration details are discussed in \citet{dolphin2012}.

The bottom panels of Figures \ref{fig:acs_sfh_models} and \ref{fig:uvis_sfh_models} include both the random and systematic errors on each of the SFHs. These inflated uncertainties have brought the SFHs from different models into formal statistical agreement, i.e., $1-\sigma$. Thus, by virtue of a more comprehensive uncertainty treatment, the SFHs are now a good representation of the true underlying SFH of the CMD.

\bsp	
\label{lastpage}
\end{document}